\documentclass{emulateapj}
\usepackage{natbib}
\usepackage{amsmath}
\usepackage{float}
\usepackage{subfigure}
\usepackage{xcolor}
\usepackage{mathrsfs}
\usepackage{hyperref}

\def\mhcm{\rm m_{H}~cm^{-3}}
 
\def\msun{M_{\odot}}

\def\cms{\rm cm^{-2}}
\def\vsc{v_{\rm s,c}}
\def\vw{v_{\rm w}}
\def\kms{\rm km~s^{-1}}
\def\rhoc{\rho_{\rm c}}
\def\rhow{\rho_{\rm w}}
\def\Mw{M_{\rm w}}
\def\dMw{\dot M_{\rm w}}
\def\dtbst{\Delta t_{\rm bst}}
 
\def\tsc{t_{\rm s,c}}
\def\tc{t_{\rm cool}}
\def\tst{t_{\rm S2}}
\def\trest{t_{\rm rest}}
\def\cv{C_{\rm v}}
\def\Rc{R_{\rm c}} 
\def\ergs{{\rm erg~s^{-1}}}
\def\ergcms{\rm erg~cm^{3}~s^{-1}}
\def\lx{L_{\rm X}}
\def\lxpeak{L_{\rm x, peak}}
\def\Tx{T_{\rm x}}
\def\be{\begin{equation}}
\def\ee{\end{equation}}
\def\nodata{--}

\shorttitle{X-ray Afterglows of TDEs}
\shortauthors{Mou et al.}

\begin{document}
\bibliographystyle{apj}
\title
{Years Delayed X-ray Afterglows of TDEs Originated from Wind-Torus Interactions }

\author {Guobin Mou\altaffilmark{1,2,3,4}, Liming Dou\altaffilmark{5,8,9}, Ning Jiang\altaffilmark{3,4}, Tinggui Wang\altaffilmark{3,4}, 
Fulai Guo\altaffilmark{6}, Wei Wang\altaffilmark{1,2}, Hu Wang\altaffilmark{1}, Xinwen Shu\altaffilmark{7}, Zhicheng He\altaffilmark{3,4}, Ruiyu Zhang\altaffilmark{6}, Luming Sun\altaffilmark{3,4}}
\altaffiltext{1}{School of Physics and Technology, Wuhan University, Wuhan 430072, China; gbmou@whu.edu.cn (GM)}
\altaffiltext{2}{WHU-NAOC Joint Center for Astronomy, Wuhan University, Wuhan 430072, China}
\altaffiltext{3}{CAS Key Laboratory for Research in Galaxies and Cosmology, Department of Astronomy, University of Science and Technology of China, Hefei 230026, China}
\altaffiltext{4}{School of Astronomy and Space Science, University of Science and Technology of China, Hefei 230026, China}
\altaffiltext{5}{
{ Department of Astronomy, Guangzhou University, Guangzhou 510006, China; doulm@gzhu.edu.cn (LD)} }
\altaffiltext{6}{Key Laboratory for Research in Galaxies and Cosmology, Shanghai Astronomical Observatory, Chinese Academy of Sciences, 80 Nandan Road, Shanghai 200030, China}
\altaffiltext{7}{Department of Physics, Anhui Normal University, Wuhu, Anhui, 241002, China}
\altaffiltext{8}{Key Laboratory for Astronomical Observation and Technology of Guangzhou, Guangzhou 510006, China}
\altaffiltext{9}{Astronomy Science and Technology Research Laboratory of Department of Education of Guangdong Province, Guangzhou 510006, China}
 
\begin{abstract} 
Tidal disruption events (TDEs) occurred in active galactic nuclei (AGNs) are a special class of sources with outstanding scientific significance. 
TDEs can generate ultrafast winds, which should almost inevitably collide with the preexisting AGN dusty tori. We perform analytical calculations and simulations on the wind-torus interactions and find such a process can generate considerable X-ray afterglow radiation several years or decades later after the TDE outburst. This provides a new origin for the years delayed X-rays in TDEs. The X-ray luminosity can reach $10^{41-42}~\ergs$, and the light curve characteristics depend on the parameters of winds and tori. We apply the model to two TDE candidates, and provide lower limits on the masses of the disrupted stars, as well as rigorous constraints on the gas densities of tori. Our results suggest that the observations of the time delay, spectral shape, luminosity and the light curve of the X-ray afterglow can be used to constrain the physical parameters of both TDE winds and tori, including the wind velocity, wind density
, cloud density and cloud size.   
\end{abstract} 


\section{Introduction}
Active galactic nuclei (AGNs) are generally considered to have tori consisting of dusty clouds (\citealt{antonucci1993, elitzur2012, netzer2015}). 
There are several methods of studying the clumpy torus, including hydrodynamic simulations (\citealt{wada2009}), geometrical models by fitting IR spectra (\citealt{nenkova2008, stalevski2012}), observations (including by coronal lines, \citealt{rose2015}; X-ray ellipses, \citealt{risaliti2002, rivers2011, markowitz2014}; or water maser in individual sources, \citealt{kondratko2005}). As key parameters, the density and size of clouds making up the torus derived from different methods vary greatly. The parameters are crucial for understanding the physics and formation of the torus, and the connection with accretion disk.   

When a star occasionally plunges into the tidal radius of supermassive back hole (SMBH), the star will be disrupted and give rise to a tidal disruption event (TDE; \citealt{hills1975, rees1988, kochanek1994}). However, the observed temperature ($\sim 10^4$ K) and bolometric luminosity (typically $10^{43}-10^{44}~\ergs$) are usually much lower than theoretical predictions, while the inferred blackbody emission radius is surprisingly large (e.g., \citealt{van2019, van2020, piran2015}). Therefore, the mass of the disrupted star is difficult to constrain by observations due to the large uncertainty in radiation efficiency (see \citealt{mockler2019} for rapid circularization and \citealt{ryu2020} for slow circularization). 
For TDEs, theoretically ultrafast outflows can be generated in two processes: the self-crossing process due to strong relativistic precessing (\citealt{lu2020, sadowski2016}) and the super-Eddington accretion (\citealt{dai2018, curd2019}).  
In the former case, the self-crossing shock can dissipate the kinetic energy, and heats up part of the materials to form outflows. Depending on the orbital parameters, the kinetic energy of outflows can reach up to $\sim 10^{52}$ erg and the outflowing mass can be a few tens percent of the disrupted star (\citealt{lu2020}) or even higher (\citealt{metzger2016}). 
In the latter case, after the debris settles into an accretion disk, simulations under various parameters have shown that energetic outflows are generated during the earlier super-Eddington accretion phase, with kinetic power of $10^{44-45}~\ergs$ (\citealt{curd2019}), or even up to $10^{46}~\ergs$ (\citealt{dai2018}). The total kinetic energy of outflow is expected to be $10^{51-52}$ erg (see also \citealt{matsumoto2020}). 
Observationally, the presence of ultrafast outflows in TDEs is confirmed directly in UV and X-ray band (\citealt{blanchard2017, blagorodnova2019, nicholl2020, hung2019}), and the high kinetic energy of winds (outflows of large opening angle) or jets (collimated outflows) has been indirectly inferred by radio emissions for some TDE candidates (e.g., \citealt{mimica2015, mattila2018, coppejans2020}). However, the direct method only reveals the physics of those components in a specific ionization state. The indirect method involves the interaction of outflow and interstellar medium, particles acceleration and synchrotron radiation of high-energy particles in an assumed magnetic field, of which the uncertainties in each step increase the uncertainty of the final results. 
Besides, the TDE wind should be transient. If the TDE wind is from the self-crossing process due to strong relativistic precession, it will last for several orbital period of the most bound debris (e.g., \citealt{sadowski2016}), while the orbital period of the most bound debris is typically less than one month. If the TDE wind is from the accretion disk, according to the high accretion rate reported in simulations (e.g., \citealt{curd2019, dai2018}), if the accretion rate is 2$\msun$/yr and the mass of the bound debris is one half of the solar mass, such a high accretion rate can only last for $\sim$3 months. Thus, we expect that the TDE wind only lasts for months.

\begin{figure*}[!htb]
   \centering
   \includegraphics[width=0.6\textwidth]{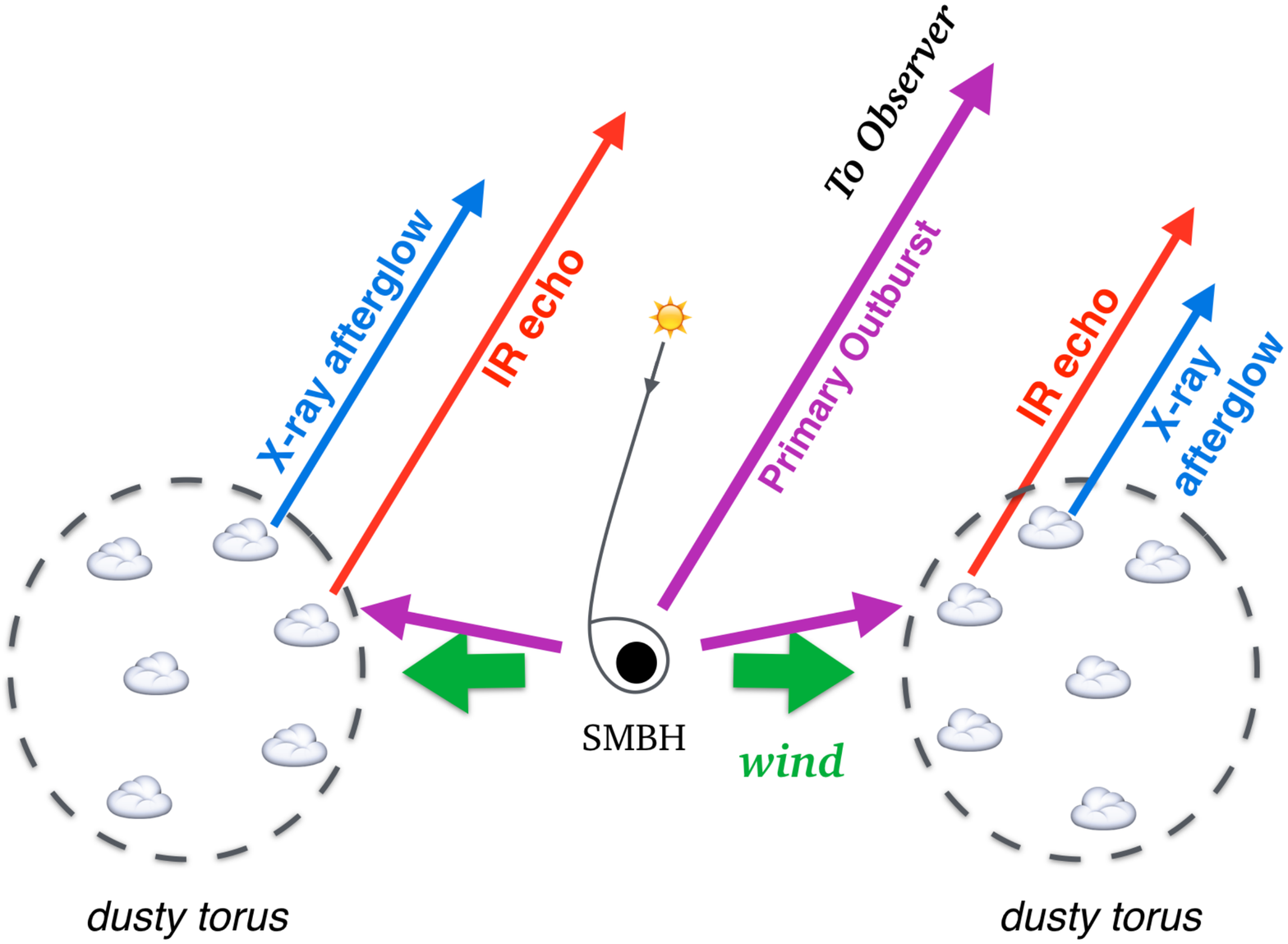} %{F2Aii.pdf} %{F2Aiii.pdf} %\vspace{1cm}
   \caption{ Schematic diagram for the X-ray afterglow. After the star is disrupted by the tidal force, winds (green arrows) can be generated during the self-crossing process of bound debris due to the relativistic precessing, or during the earlier super-Eddington accretion phase. Radiation from the TDE primary outburst (purple arrows) irradiates the surrounding dusty torus, and reprocesses in infrared emissions with time lag of tens to hundreds of days (infrared echo, red arrows). After several years or decades, the winds impact the clumpy torus, and drive shocks inside the clouds, which will heat the clouds and generate X-rays (X-ray afterglow, blue arrows). } 
   \label{plot1}
\end{figure*}

It is naturally expected that some TDEs take place in AGNs with dusty tori.
TDE wind will collide with the dusty torus almost inevitably. 
When a cloud is overtaken by a fast wind with velocity $\vw$, a shock will be driven inside the cloud with velocity of 
$\vsc=\chi^{-1/2} \vw$, where $\chi \equiv \rho_{\rm c}/\rho_{\rm w}$ is the density contrast between the cloud and wind (\citealt{mckee1975}; $\chi \gg 1$ in our concerns). The cloud materials swept up by the shock will be heated into a temperature of $k_{B} T_{\rm s,c}=\frac{3}{16} \mu m_{\rm H} v^2_{\rm s,c} \simeq 0.1 m_{\rm H} v^2_{\rm s,c}$, where $\mu$ is the mean molecular weight (0.62 for solar metallicity). 
For example, if $\vw =0.1$c and $\chi \lesssim 10^3$, the temperature of the shocked cloud can be $\gtrsim 1\times 10^7$ K, and the cloud will radiate in X-rays. 
The inner edge of the torus can be inferred by the dust sublimation radius in AGN (typically $\sim 0.1$ pc for $L_{\rm bol} \sim 10^{44-45}~\ergs$, \citealt{netzer2015}), or by the time lag of infrared echo (\citealt{dou2016,jiang2016,van2016}), or the line width of iron coronal lines arising after the outburst (e.g., \citealt{komossa2008}).   
Depending on the wind velocity and distance of the torus, the wind-torus interaction will typically occur years or decades after the TDE primary outburst in the optical/UV band and the \emph{infrared echo} (see Figure 1 for the schematic diagram). Therefore we call it the ``X-ray afterglow'', from which $\vw$ can be directly restricted from the time lag of X-rays and the radius of inner edge of dusty torus.

%%%%%%%%%%%%%%%%%%%%%%%%%%%%%%%%%%%%%%%
%%%%%%%%%%%%%%%%%%%%%%%%%%%%%%%%%%%%%%%

\section{Models and Applications}

\subsection{Energy Conversion Efficiency}
During the wind-cloud interaction, only a small fraction of the wind's kinetic energy ($E_{\rm wind}$) can be converted into cloud's energy ($E_{\rm c}$).   
The kinetic power of the wind is $0.5 \dMw \vw ^2 = 2 \pi r^2 \rhow  \vw^3$, while the internal energy gained by shocked clouds per unit time is $\frac{9}{32} \cv 4\pi r^2 \rhoc \vsc^3$ for Mach number $\gg$ 1 ($\cv$ is the covering factor). 
Therefore the energy conversion efficiency (from wind to cloud) is $\eta_{E} \equiv E_{\rm c}/E_{\rm wind} \simeq 0.56 \cv \chi^{-1/2} $. 
Without considering the energy loss in adiabatic expansion, this indicates that the total radiation energy in the X-ray afterglow ($\simeq E_{\rm c}$) is determined by the density contrast and the total wind energy $E_{\rm wind}$: 
\be
E_{\rm c} = \eta_{E} E_{\rm wind} \simeq 1.8 \% ~ \cv \chi^{-0.5}_3 E_{\rm wind}
\ee
where $\chi_3 \equiv \chi/10^3$.  
$E_{\rm c}$, $\vw$ and $r$ ($\sim R_{\rm sub}$) are observable parameters, and $\cv$ is on the order of $10^{-1}$. The conversion efficiency relationship is also verified by simulation tests in appendix.  
According to this relationship, the total X-ray energy during an X-ray afterglow can reach a considerable value of several percentage of the wind's kinetic energy if the density contrast is not too high. Besides, it is worth noting that essentially $\cv$ in Equation 1 is the fraction of the outflow (winds or jets) energy blocked by the torus. For non-isotropic wide-angle winds or beamed jets, $\cv$ is not necessarily equal to the torus covering factor.

\begin{figure*}[!htb]
   \centering
   \includegraphics[width=0.8\textwidth]{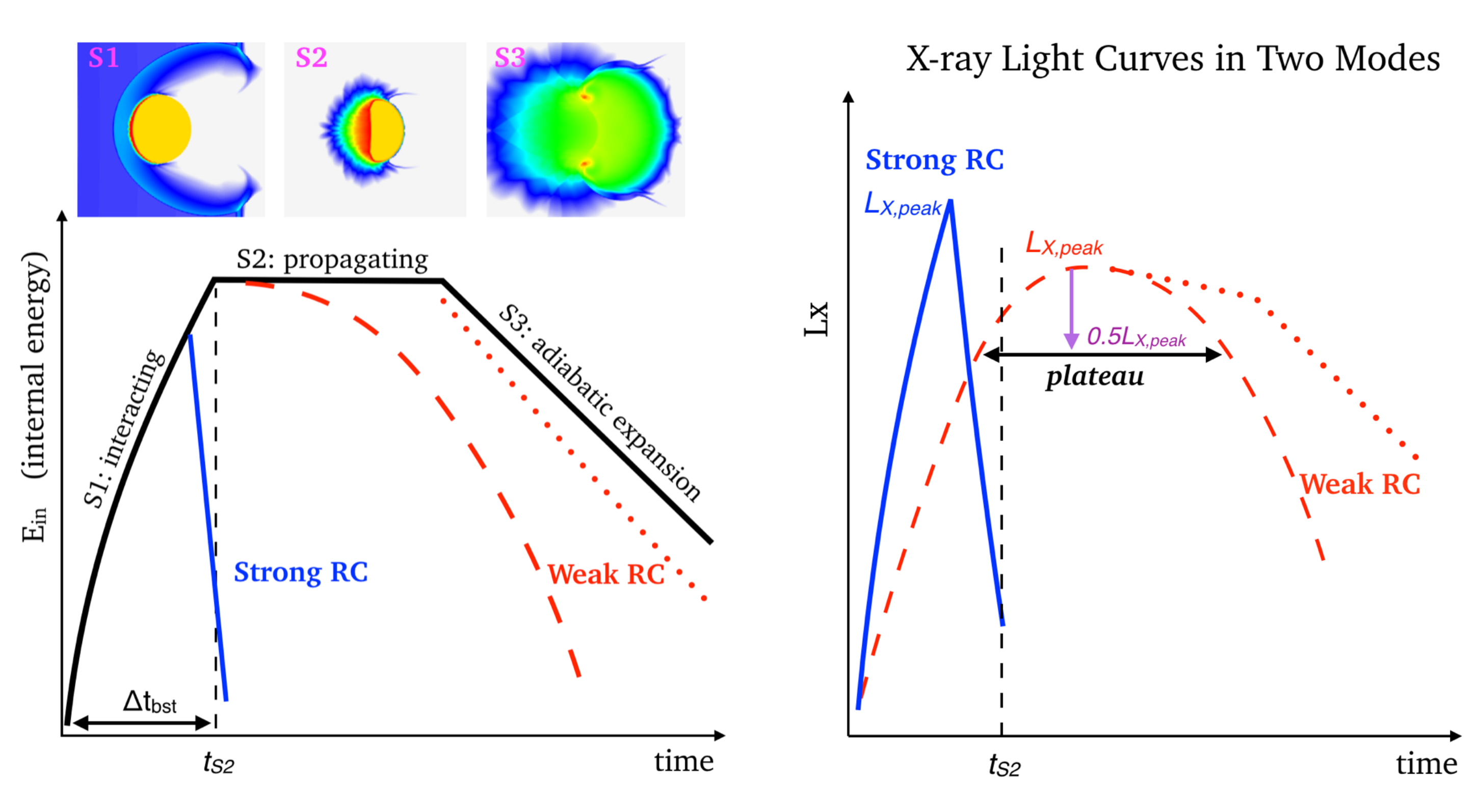} 
   \caption{Theoretical model for the wind+one cloud instantaneous interaction. 
Left panel shows the evolution of the total internal energy of the shocked cloud over time in the adiabatic case (solid black line) and in the radiative cooling (RC) case (solid blue -- strong RC, dashed red -- weak RC, dotted red -- weak RC when $\tc > \tsc$). In the adiabatic case, the internal energy increases rapidly during S1 (wind-cloud interacting stage), and then roughly keeps unchanged during S2 after the wind shuts down (cloud shock propagating stage). After swept over by the cloud shock, the cloud undergoes an adiabatic expansion process (S3). 
Radiative cooling affects the evolution of the internal energy depending on the radiative cooling timescale.   
The X-ray light curves show two modes (right panel): strong RC mode (blue solid line) and weak RC mode (red dashed/dotted lines). 
For strong RC, the X-ray light curve varies drastically with time. 
For weak RC, the luminosity evolves slowly and shows a ``plateau'' during S2. The subsequent decline trend depends on the cooling process of shocked cloud.  (See appendix for simulation tests.)  }
   \label{plot2}
\end{figure*}

\subsection{the Process of Wind-Cloud Instantaneous Interaction}

The model of wind and cloud interaction has been applied in supernova remnants (e.g., \citealt{mckee1975, chugai1994}), wherein the wind (or more accurately, supernova ejecta) is treated as steady state.  
However for TDEs, the strong wind only lasts for months, since the strong self-crossing of stellar debris is a short-term process, and alternatively, the high accretion mode is short-lived. 
We assume that the wind only lasts for $\dtbst$, and after $\dtbst$ it ``shuts down'' at the origin, while the wind that has been generated continues to travel outwards. 
For a steady wind-cloud interaction, the cloud shock sweeps across the entire cloud with a timescale of $2 \Rc/\vsc$ = 6 yr $(\Rc /10^{16} {\rm cm}) (\vsc/1000 {\rm km s^{-1}})^{-1}$. For those cloud sizes of $\gtrsim 1\times10^{-3}$ pc (\citealt{kondratko2005, nenkova2008, stalevski2012}), the cloud shock probably has not swept across the entire cloud when the TDE wind shuts down.  
Therefore the wind-cloud interaction concerned here is an instantaneous process, and by performing simulation tests (Figure 7 in appendix), we conclude that this process can be divided into three stages (Figure 2). 
\footnote{We caution here that the clouds in the dusty torus should be strip-shaped stretching along the rotation direction due to the tidal shearing, which is also present in hydrodynamic simulations for torus (\citealt{wada2009}). Our following discussions apply to clouds of this shape. } 

S1, wind-cloud interacting stage at $0 < t \leqslant \dtbst =\tst$ ($t=0$ is the start time of the interaction). 
During this stage, as the cloud shock is sweeping up the cloud, more and more wind's kinetic energy is converted into the internal energy of clouds. Along with this process, X-ray emission arises.

S2, cloud shock propagating stage at $\tst < t < \tst+\tsc$ after the wind shuts down (the duration $\tsc$ is the timescale of the shock sweeping across the cloud). The cloud shock inside has not yet swept over the entire cloud during this stage, and both radiative cooling and adiabatic cooling may affect the radiation. 
However, interestingly we find that the total internal energy of the shocked cloud $E_{\rm c}$ roughly remains unchanged until the shock sweeps across the entire cloud.  
Therefore during S2, we can assume that the internal energy of the cloud can only be lost through radiative cooling, with a timescale of $\tc = (\gamma -1)^{-1} (n_e+n_I) k_{\rm B}T_{\rm s,c}/\Lambda n_e n_H \sim 1.8 {\rm yr} ~ T_{7} \Lambda^{-1}_{-23} \rho^{-1}_{7}$ where $\gamma$ is the adiabatic index, $\Lambda$ is the cooling function 
 normalized by $10^{-23}\ergs$cm$^{3}$
(\citealt{sutherland1993}), 
$T_7 \equiv T/10^7$K, and $n_e$, $n_I$ and $\rho_7$ are the density values of the shock cloud ($\rho_{7} \equiv \rho_{\rm c}/10^7 \mhcm$). 

S3, adiabatic expansion stage after the cloud shock has swept over the entire cloud at $t \geqslant \tst+\tsc$.
In adiabatic process without radiative cooling, we find that the internal energy of shocked cloud $E_{\rm c}$ decreases as a power-law function of time: $E_{\rm c} (t) = E_{\rm c} (\tst) \left( \frac{t-\tst}{\tsc} \right)^{-m} $, 
where $E_{\rm c}(\tst)$ is the internal energy of the cloud at $t=\tst$ and is roughly a constant during S2. 
Simulation tests indicate that $m \simeq 1.4$, which is consistent with the exponent of $4/3$ given by a ring-shaped cloud expanding at a constant speed.

\subsection{X-rays from Wind-Cloud Instantaneous Interaction}

When radiative cooling is included, depending on the radiative cooling timescale, the X-ray luminosity can be estimated in two cases or modes (Figure 2). 

1. Strong radiative cooling (strong RC) mode: $\tc \lesssim \dtbst$.   
In this case, the cooling is so strong that the internal energy of the shocked cloud gained per unit time is rapidly lost in radiation. 
Therefore the typical X-ray luminosity is 
\be
\centering
\begin{split} 
\lx  \sim E_{\rm c}(t)/t = \eta_{E} ~ \dMw \vw^2 /2   ~~~~~~~~~~~~~~~~~ \\
 ~~~ = 5 \times 10^{41} ~\ergs \cv  r^{-1}_{0.1} \rho^{-0.5}_{\rm c,7} \dot{m}^{1.5}_{\rm w} v^{1.5}_{\rm w,4} ~,
\end{split}
\ee
in which $\dot{M}_{\rm w}$ is the mass outflow rate of the wind, $\dot{m}_{\rm w} \equiv \dot{M}_{\rm w}/1 M_{\odot} {\rm yr^{-1}}$, $v_{\rm w,4}\equiv \vw/10^4 ~\kms$, $r_{0.1}\equiv r/0.1$pc and $r$ is the distance from the inner edge of the torus to the SMBH. 
The strong RC mode generally corresponds to the case of high-density cloud. 
By equating $\tc$ and $\dtbst$, we obtain a critical density of $\rhoc \simeq 2\times 10^8 \mhcm T_7 (\Delta t_{\rm bst,month} \Lambda_{-23})^{-1}$ ($\Delta t_{\rm bst,month} \equiv \dtbst/1$month, $T_7\equiv T/10^7K$, $\Lambda_{-23}\equiv \Lambda/10^{-23}\ergs$cm$^{-3}$), above which the X-ray radiation is in strong RC regime.
The X-ray light curve behaves like dramatic flares with rapid rise and decline, which is the signature of this mode. 

2. Weak RC mode: $\tc \gtrsim \dtbst $.  
In this case the total energy of the cloud gained during S1 will be lost in radiative cooling during S2 and radiative + adiabatic cooling during S3.
The X-ray luminosity in S2 can be estimated by: 
\be
\centering
\begin{split}
\lx \sim E_{\rm c}/\tc =2\cv \Lambda \rhoc \chi^{0.5} \Mw /m^2_{\rm H}  ~~~~~~~~~~~~ \\
~~ \simeq 4 \times 10^{42} ~\ergs \cv \Lambda_{-23} \rho_{\rm c,7} \chi^{0.5}_3  m_{\rm w}  \\
~~ \simeq 1 \times 10^{42} ~\ergs  \cv \Lambda_{-23} r^2_{0.1} v_{\rm w,4} \rho^{1.5}_{\rm c,7} \rho^{0.5}_{\rm w,4}
\end{split}
\ee
in which $\Mw=\dMw \dtbst$ is the wind mass, $m_{\rm w} \equiv \Mw/M_{\odot}$, $\rho_{\rm w,4} \equiv \rhow/10^4 \mhcm$. 
During S2 the X-ray light curve shows a plateau with timescale of $\sim \min(\tc,\tsc)$ (Figure 2), since the X-ray emissivity $\epsilon(T)$ in $\lx = \int \epsilon(T) n_e n_H dV$ does not vary significantly within a large temperature range (Figure 8 in appendix). 
During S3, the decline rate of the X-ray luminosity depends on the radiative cooling timescale. When $\tc$ is so large that cooling is dominated by adiabatic expansion, the X-ray luminosity will be modulated by adiabatic cooling.
In this case, since $E_{\rm c} (t) \propto  (t-\tst)^{-m}$ and $pV^{\gamma} \simeq$ constant for adiabatic approximation, we have $\rho(t) \propto (t-\tst)^{-1.5m}$, and the X-ray luminosity in S3 thus decreases in a power-law form of time: $\lx \propto (t-\tst)^{-1.5m}$ with the index $-1.5m \simeq -2.1$ (Figure 7 in appendix). 

According to above analysis, the typical X-ray luminosity in X-ray afterglow is $\sim 10^{41}~\ergs$, and under certain conditions, the luminosity can be up to a few times $10^{42}~\ergs$, or even higher. 

%%%%%%%%%%%%%%%%%%%%%%%%%%%%%%%%%%%%%%%
%\subsection{Applications to TDE Candidates for the TDE X-ray Afterglows by Numerical Simulations}
%\subsection{Possible Candidates for the TDE X-ray Afterglows}
\subsection{Applications to Candidates}
As applications of our model, here we selected two TDE candidates: 
SDSS J095209.56+214313.3 (hereafter J0952) and PS1-10adi. J0952 is a TDE candidate with strong transient coronal lines (\citealt{komossa2008, komossa2009, wang2012, yang2013}), and UV flare (\citealt{palaversa2016}). PS1-10adi is a highly energetic transient event  (\citealt{kankare2017}). Interestingly, strong dust echoes are discovered in both two objects (\citealt{dou2016, jiang2019}). The rising and plateau in hard X-ray for two years in J0952, and the UV-optical rebrightening and contemporaneous X-ray onset in PS1-10adi detected 3-5 years after the optical peak (\citealt{auchettl2017, jiang2019}) could not be well explained by the classical TDE scenario. 
It is uncertain whether J0952 harbors a pre-existing AGN before TDE outburst, however the emergence of strong iron coronal lines with TDE (\citealt{komossa2008}) suggests that there should be a torus. PS1-10adi is clearly a Type 1 AGN on basis of its post-flare optical spectra with persistent broad lines, although there is no available spectral data before TDE. 
Thus, the X-ray emission feature may be explained by the interaction between the high-velocity wind and the torus (\citealt{jiang2019}). 

We made the data reduction and obtained their X-ray spectra  (see the Appendix for details). The observed fluxes we obtained are consistent with past studies (\citealt{auchettl2017, kankare2017}). Both spectral profiles are flat, which may indicate the sources are (partially) obscured. 
For J0952, three X-ray data points sketch the rising and declining phases. 
For PS1-10adi, the X-ray luminosity is as high as a few $10^{42}\ergs$ to $10^{43}~\ergs$, which is close to the upper limit in our model, and can be used to explore the conditions required to form relatively ``extreme'' X-ray afterglow phenomena.

Interactions of wind and torus are complex. As a first-order approximation, the wind-torus interactions can be regarded as the sum of a series of wind + single cloud interactions. However, after passing the bow shock ahead of the clouds, the wind will significantly slow down and the interaction will be significantly weakened. Therefore, the model's restrictions on the cloud's parameters only apply to those clouds located at the inner edge of the torus. 
For more accurate results, interactions of wind + multi-clouds should be resorted to hydrodynamic simulations. 
The hydrodynamic equations are as follows:  
\begin{gather}
  \frac{d \rho}{d t} + \rho \nabla \cdot {\bf v} = 0,\label{hydro1} \\
 \rho \frac{\partial \bf v}{\partial t} +\rho {\bf v} \cdot \nabla {\bf v} = -\nabla p , \\
   \frac{\partial e}{\partial t} +\nabla \cdot(e{\bf v})=-p \nabla \cdot {\bf v}+\mathcal{L}_{c}, \label{hydro3}  \\
   \mathcal{L}_{c}= -n_{e} n_{i} \Lambda_{N} , 
\end{gather}
in which $p=(\gamma-1)e$ is the thermal pressure ($\gamma = 5/3$ is the adiabatic index for ideal gas), 
$n_e$ and $n_{i}$ represent the number densities of electrons and all species of ions, $\Lambda_{N}$ is the cooling function for which we fit the collisional ionization equilibrium data for solar abundance (\citealt{sutherland1993}) as $\Lambda_{N}=2.2\times 10^{-27}T^{0.5} + 2.0\times 10^{-15} T^{-1.2}+2.5 \times 10^{-24}$ for $T\geqslant 1\times 10^5$ K, and $2.0\times 10^{-31}T^{2.0}$ for $T< 1\times 10^{5}$ K ($\Lambda_N$ in units of $\ergcms$ and $T$ in units of Kelvin). Here due to the timescale we concern is very short, the gravity of the SMBH and the rotation of the cloud can be neglected.  

We adopt the APEC (\citealt{smith2001}) program to calculate the X-ray luminosity and spectrum in simulations. We assume that the hot gas is optically thin and under collisional ionization equilibrium, and the gas metallicity is $Z=Z_{\odot}$. The X-ray luminosity in 0.3-10 keV (corresponding to the energy range of Swift XRT) can be calculated as follows: 
\begin{equation}\label{brightness}
L_{\rm X}= { \int n_{\rm e}n_{\rm H}\epsilon(T) dV},
\end{equation}
where $\epsilon(T)$ is the X-ray emissivity of the hot gas in 0.3--10 keV band in units of erg cm$^3$ s$^{-1}$.
We fit $\epsilon(T)$ data by the following forms (note here $\Tx \equiv  \log T(K)$) , and plot both the fitting curve and the data in Figure 8 in appendix: 

(1), for $T_{\rm x} < 6.7$,
\be
\epsilon(T)= 10^{-4(\Tx -6.45)^{2.0}-22.7}+10^{-5(\Tx -6.9)^{2.0}-22.65} , \\
\ee

(2), for $\Tx \geqslant 6.7$, 
\be
\begin{split}
\epsilon(T) & = 10^{-4(\Tx -6.45)^{2.0}-22.7}+10^{-5(\Tx -6.9)^{2.0}-22.65}  \\
  & +10^{-0.25(\Tx -8.15)^{2.0}-22.68} . 
\end{split}
\ee

Spectrum energy distribution in X-ray band is determined by:  
\begin{equation}\label{sed}
F_{\rm E}= {\int_{V} n_{\rm e}n_{\rm H} f_{E}(T) dV}= \sum_{T_i} f_{E}(T_i) W(T_i) , 
\end{equation}
in which $F_{\rm E}$ is in units of photons s$^{-1}$ cm$^{-2}$ keV$^{-1}$, $E$ is the energy of X-ray photons, $T$ is the gas temperature, $f_{E}(T)$ is the spectrum from collisionally-ionized diffuse gas with single temperature which is output from APEC based on the AtomDB atomic database, $dV$ is the grid volume in our hydrodynamic simulations, and the weight is $W(T_i) = {\int _{(T_i-0.5\Delta T_{i-1}) < T < (T_i+0.5\Delta T_i)}  n_{\rm e}n_{\rm H} dV}$. We assign the temperature $T_i$ in geometric sequence from 0.279 keV to 22.186 keV with a common ratio of 1.2: 0.279 keV, ...., 1.00 keV, 1.20 keV, ...., 22.186 keV. 
For simplicity, we do not consider the absorption of X-rays by clouds, nor the time lags due to the different distances from the clouds to the observer.

\begin{table*}
  \centering
%   \begin{minipage}{120mm}
  \renewcommand{\thefootnote}{\thempfootnote}
  \caption{Parameters of simulation domains for models.  (1) model label, (2) the inner and outer boundary in radial direction, (3) the range of $\theta$, (4) number of grids in $r-$direction and $\theta-$direction ($N_r$, $N_\theta$), (5) $R_{\rm torus}$ -- the radius of the hypothetical circular torus profile (blue circle in Figure 3), (6) $r_{\rm torus}$ -- distance from the inner edge of the torus to the SMBH, (7) total number of clouds inside the hypothetical torus profile. }
  \begin{tabular}{@{}  c  c  c  c  c  c  c}
  \hline
         Name
        & ($r_{\rm in}$, $r_{\rm out}$)
        & $\theta$
        &  ($N_r$, ~ $N_\theta$) 
        & $R_{\rm torus}$
        & $r_{\rm torus}$
        & $N_{\rm c}$
        \\
    \hline
     J1,J2Dw, J5Dc--J7Dc     & (0.05635 pc, 0.177 pc)  & $39^{\circ}-141^{\circ}$ & (5200, 6600) & 0.20 pc & 0.06 pc & 213   \\ % 
    J3Dc  & (0.05635 pc, 0.2185 pc)  & $39^{\circ}-141^{\circ}$ & (6160, 6600) & 0.20 pc & 0.06 pc & 213   \\ %
    J4Rc  & (0.05635 pc, 0.177 pc)  & $39^{\circ}-141^{\circ}$ & (8640, 11880) & 0.20 pc & 0.06 pc & 3299 \\ % SE57 or SJ51show
    PS  & (0.07742 pc, 0.267 pc) & $43^{\circ}-137^{\circ}$ & (6880, 7200) & 0.20 pc & 0.08 pc & 370 \\ %
    \hline
 \label{table1}
  \end{tabular}
%  \end{minipage}
\end{table*}

\begin{table*}
  \centering
  \renewcommand{\thefootnote}{\thempfootnote}
  \caption{Model parameters and results for J0952 (model J1--J5Rs) and PS1-10adi (model PS): (1) model label, (2) density of clouds, (3) radius of clouds, (4) density of winds normalized at $r=0.1$ pc, (5) wind speed over the light speed, (6) peak X-ray luminosity, (7) duration of the X-ray plateau (above $0.5\times \lxpeak$), (8) total X-ray radiation energy, (9) total wind mass. }   
  \begin{tabular}{@{} c  c  c  c  c  c  c  c  c} 
  \hline
        & {$\rhoc$}
        & {$\Rc$ }
        & {$\rhow(0.1)$}
        & {$\vw /c$} \footnote{ The wind velocity is higher than $r_{\rm torus}$ over the observed X-ray time lag, since the observed X-ray data listed in our article (see appendix for details) are not obtained from the high cadence observations, thereby missing the rise stage of the X-rays afterglow. } 
        & {$\lxpeak$ }
        & {$t_{\rm plat}$ }
        & {$E_{\rm X}$}
        & {$m_{\rm w}$}
        \\
    Run & ($\mhcm$)  & (pc) & ($\mhcm$) &   & ($\ergs$)  & (yr)     &  erg & $\msun$ \\
    (1)   &  (2)        &     (3)      & (4)           &  (5)            &  (6)  &  (7)  & (8) & (9)  \\
    \hline
  J1     & $4.0 \times 10^6$ & $1.6\times 10^{-3}$ & $6.0\times 10^4$ & 0.08 & $8.2\times 10^{41}$ & 1.8 & $5.3\times 10^{49}$ & 0.75 \\ % se55ST
 J2Dw & $4.0 \times 10^6$ & $1.6\times 10^{-3}$ & $1.5\times 10^4$ & 0.08 & $3.2 \times 10^{41}$ & 1.9 & $1.9\times 10^{49}$ & 0.19 \\ % se56d
 J3Dc & $2.0 \times 10^6$  & $1.6\times 10^{-3}$ & $6.0\times 10^4$ & 0.08 & $2.8\times 10^{41}$ & 2.4 & $2.7\times 10^{49}$ & 0.75 \\ % se60D
 J4Rc & $4.0 \times 10^6$  & $0.4\times 10^{-3}$ & $6.0\times 10^4$ & 0.08 & $3.1\times 10^{41}$ & 1.4 & $1.7\times 10^{49}$ & 0.75 \\ % se93R04
 J5Dc & $2.0 \times 10^7$  & $1.6\times 10^{-3}$ & $6.0\times 10^4$ & 0.08 & $1.9\times 10^{42}$ & 0.5 & $7.5\times 10^{49}$ & 0.75 \\ % se59D
 J6Dc & $8.0 \times 10^7$  & $1.6\times 10^{-3}$ & $6.0\times 10^4$ & 0.08 & $1.1\times 10^{42}$ & 0.2\footnote{
 $t_{\rm plat}$ is the largest value among the flares for J6Dc (also the case for J7Dc). } & $1.9\times 10^{49}$ & 0.75 \\ % se65D87
 J7Dc & $3.2 \times 10^8$  & $1.6\times 10^{-3}$ & $6.0\times 10^4$ & 0.08 & $2.9\times 10^{41}$ & 0.1 & $3.9\times 10^{48}$ & 0.75 \\ % se67D38
   PS  &  $1.3 \times 10^7$ & $1.6\times 10^{-3}$ & $7.6\times 10^4$ & 0.11 & $8.4\times 10^{42}$ & 1.2  & $3.6\times 10^{50}$ & 1.31  \\ % P61 
    \hline
 \label{table2}
  \end{tabular}
\end{table*}

We adopt the ZEUSMP code to solve the above equations (\citealt{stone1992, hayes2006}), and choose 2.5D Spherical coordinates in which $\phi$-direction is set to be symmetric. 
The inner edge is $r_{\rm torus} \sim 0.06$ pc for J0952 (inferred by the line width of the transient iron coronal lines; \citealt{komossa2008}) and 0.08 pc for PS1-10adi (inferred by the time lag of infrared echo and the dust sublimation radius; \citealt{jiang2019}). 
The size of the cloud is poorly understood in the present. So far, the few observations are given by X-ray eclipse events or water maser, in which the size of the cloud is suggested to be a few $10^{15}$cm (\citealt{rivers2011}),  $10^{14-16}$cm (\citealt{markowitz2014}, including broad line region clouds therein), or $0.001-0.006$ pc (\citealt{kondratko2005}).  In our simulations, 
the clumpy torus is simplified as hundreds of circular clouds with radius of $5\times10^{15}$cm (0.0016 pc), which are randomly distributed in $r$ and $\theta$ directions within a hypothetical torus border which is a large circle with radius of $R_{\rm torus}$ (Table 1 and Figure 3).  
We also investigate the case of smaller clouds with $\Rc=1.25\times 10^{15}$cm (model J4Rc). 
The TDE wind is injected isotropically at the inner boundary $r_{\rm in}$ (Table 1) with velocity of $\vw$ and density of $\rhow(r_{\rm in})=\rhow (0.1) (0.1{\rm pc}/r_{\rm in})^2$, and the duration of the wind $\dtbst$ is fixed in 2 months for all runs. The parameters of $\rhoc$, $\Rc$, $\rhow(0.1)$ and $\vw$ are listed in Table 2.  

The parameters of simulation domains are listed in Table 1. The computation domain is divided into thousands of non-uniform pieces in $r-$direction with $dr_{i+1}/dr_{i}=$constant 
in which the ratio is 1.00022 for J0952 (the ratio in model J4Rc is 1.000132) and 1.00018 for PS1-10adi, 
and uniform pieces in $\theta-$direction. The high resolution adopted in our simulations well resolves each cloud, and ensures that the results are reliable (see appendix for resolution tests).

\section{Results}

Here we only present one model for PS1-10adi due to less X-ray data, and we present several models for J0952 to explore the influence of the parameters on the results. 
The best-fit models are \emph{J1} for J0952 and \emph{PS} for PS1-10adi. 
We plot the snapshots of density, temperature and the X-ray volume emissivity distributions for model J1 and model PS in Figure 3. The cloud materials swept by the cloud shock are heated to $10^7 - 10^8 K$, and give rise to X-ray afterglows. The X-rays are mainly contributed by shocked cloud due to the high density, especially those materials just swept by the shock. The X-ray volume emissivity can reach $10^{-9} - 10^{-8}$ erg cm$^{-3}$ s$^{-1}$.

Together with observational data, 
we plot the simulated X-ray light curves in Figure 4, and the X-ray spectrum of J0952 in Figure 5. 
As expected, the interactions of strong TDE wind and the clumpy torus can indeed generate X-ray afterglow years after the optical outburst, with X-ray peak luminosity of $10^{41-42}~\ergs$.  
Comparing J1 and J2Dw, when the wind density drops to 1/4 of J1 case, the peak X-ray luminosity will decrease to $\sim$ 40\% of J1's value. 
Comparing J1 and J3Dc, when the cloud density drops to one half, the peak X-ray luminosity will decrease to 34\% of J1's value. Those are roughly consistent with the law in Equation 3. 
Comparing J1 and J4Rc, when we decrease the cloud size while keeping the total cloud mass roughly unchanged by increasing the number of clouds (Table 1), we find that the rising stage of X-ray luminosity is significantly shortened and the peak luminosity is lower. After passing the peak value, the X-ray luminosity soon turns into a power law decline: $\lx \propto (t-2.5)^{-2.1}$, where $t$ is in units of year, and 2.5 yr is the moment when X-ray afterglow begins. This is due to the fact that when the cloud size is small, lots of post-shocked clouds would enter stage S3 soon, and expand before radiative cooling.  
When the density is $\sim 2\times 10^7~\mhcm$ (model J5Dc), it can be regarded as a transition case between strong RC and weak RC mode. If only examining the dependence on the cloud density, the X-ray luminosity reaches the maximum value in the transition case. 
When the density is $\gtrsim 1\times 10^8~\mhcm$, the X-ray radiation enters the strong RC regime, in which the luminosity evolves dramatically in a timescale of one month (J6Dc) to a few days (J7Dc), and the X-ray spectrum is softer. From J6Dc and J7Dc, we can also find that when the cloud density is higher, both the peak X-ray luminosity and the total X-ray radiation energy are lower. 
The X-ray luminosity in strong RC regime seems to be much lower than the prediction of equation 2, which is due to the ``effective'' covering factor cannot take the global value of torus, but is significantly reduced and approaches the covering factor of one single cloud due to the short cooling timescale.  
Thus, from the hydrodynamic results on the wind-torus interactions, we conclude that: 
1) the wind density affects X-ray luminosity roughly as expected in equation 3; 
2) cloud size can affect the X-ray light curve via fast (small size) or slowly (large size) entering S3 stage for the cloud, and when the cloud size is smaller, the X-ray luminosity reaches its peak value and turns into the power-law decay sooner; 
3) the variability in X-rays and the luminosity are sensitive to the cloud density.

According to our best-fit models, the cloud density
required to generate observed X-rays should be a few $\times 10^6 \mhcm$ for J0952, and $\sim 1\times 10^7~\mhcm$ for PS1-10adi. 
This is consistent with the density range inferred from the coronal lines (\citealt{rose2015, yan2019, wang2012}), 
 and X-ray eclipse event (\citealt{rivers2011}). 
%But more importantly, those coronal lines can only indicate the existence of gas at a certain density range without telling the ``typical'' density value, while our results further reveal the typical density value of the torus. 

 Our results suggest that, 
the total wind mass is $0.75 \msun$ for J0952, and $1.31 \msun$ for PS1-10adi. 
Considering that the wind materials come from one half of the disrupted star, the mass of the initial star should be at least 1.5 and 2.6 $\msun$ for J0952 and PS1-10adi, respectively. 
Both of the inferred wind masses are noticeably around one solar mass, of which the coincidence implies that these two transients should indeed be TDEs.  
The wind's kinetic luminosity is $0.83\times 10^{45}~\ergs$ for J0952, and $2.7\times 10^{45}~\ergs$ for PS1-10adi, which are very energetic compared with the Eddington luminosity of a $10^7\msun$-mass SMBH.  
The total kinetic energy is $4.3\times 10^{51}$ erg for J0952 and $1.4\times 10^{52}$ erg for PS1-10adi. 
Therefore the total X-ray radiation (Table 2) from the afterglow is 1-2\% of the total kinetic energy of winds, as expected in above analysis.

%%%%%%%%%%%%%%%%%%%%%%%%%%%%%%%%%%%
%%%%%%%%%%%%%%%%%%%%%%%%%%%%%%%%%%%

\begin{figure*} %[!htb]
   \centering
   \includegraphics[width=0.70\textwidth]{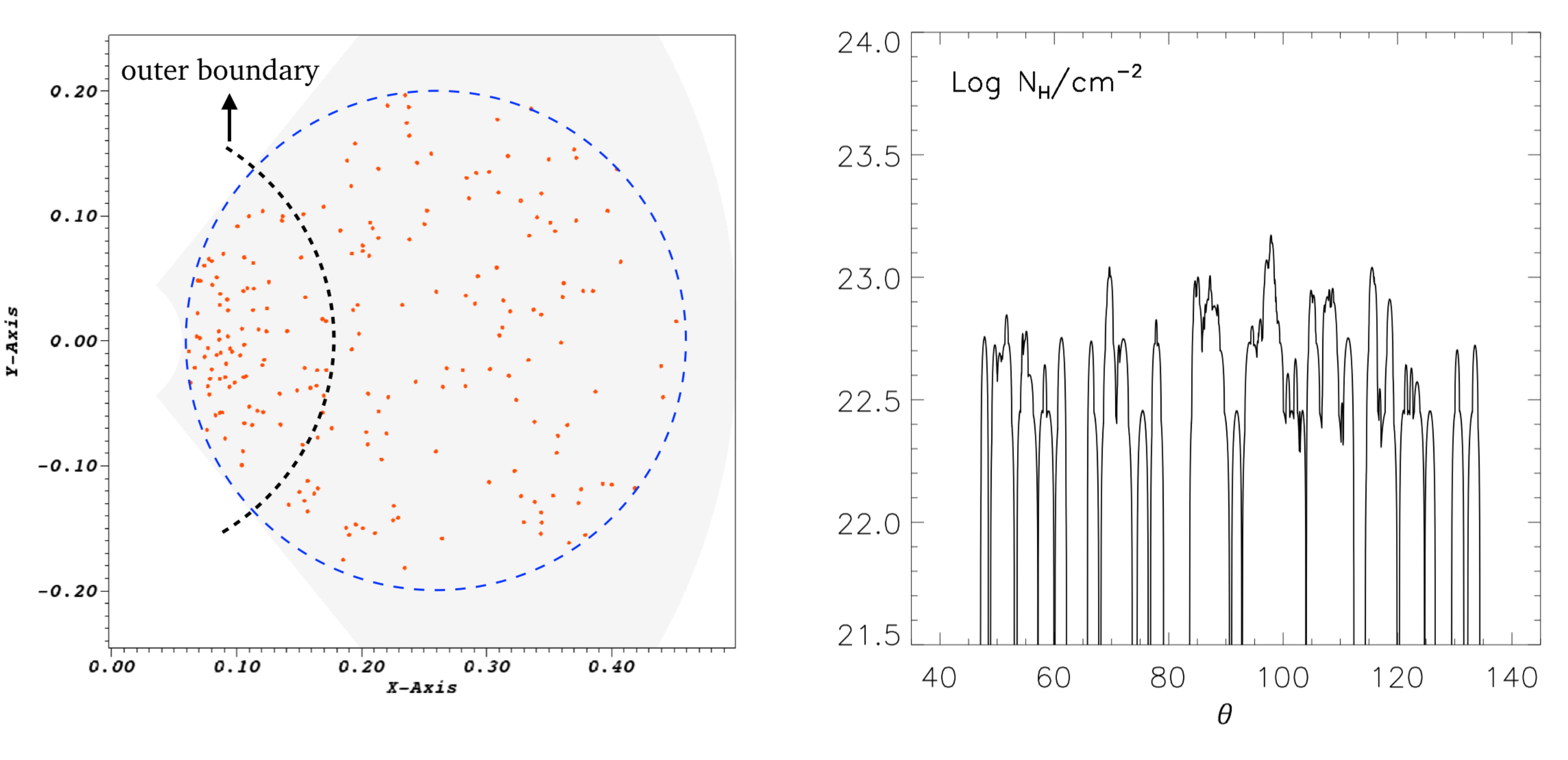}  %{SP3A.pdf} 
   \includegraphics[width=0.70\textwidth]{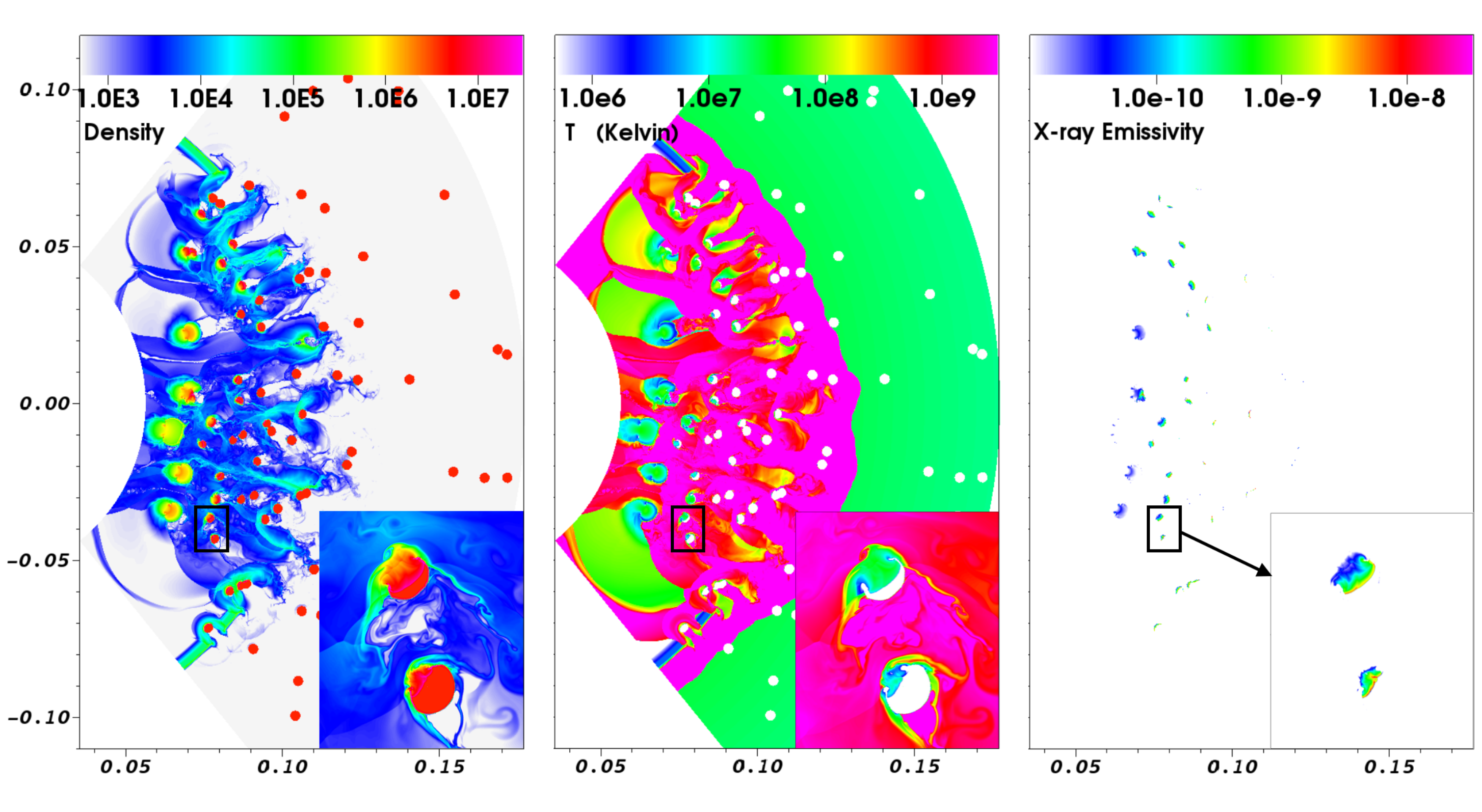}  %{SP3B.pdf} 
   \includegraphics[width=0.72\textwidth]{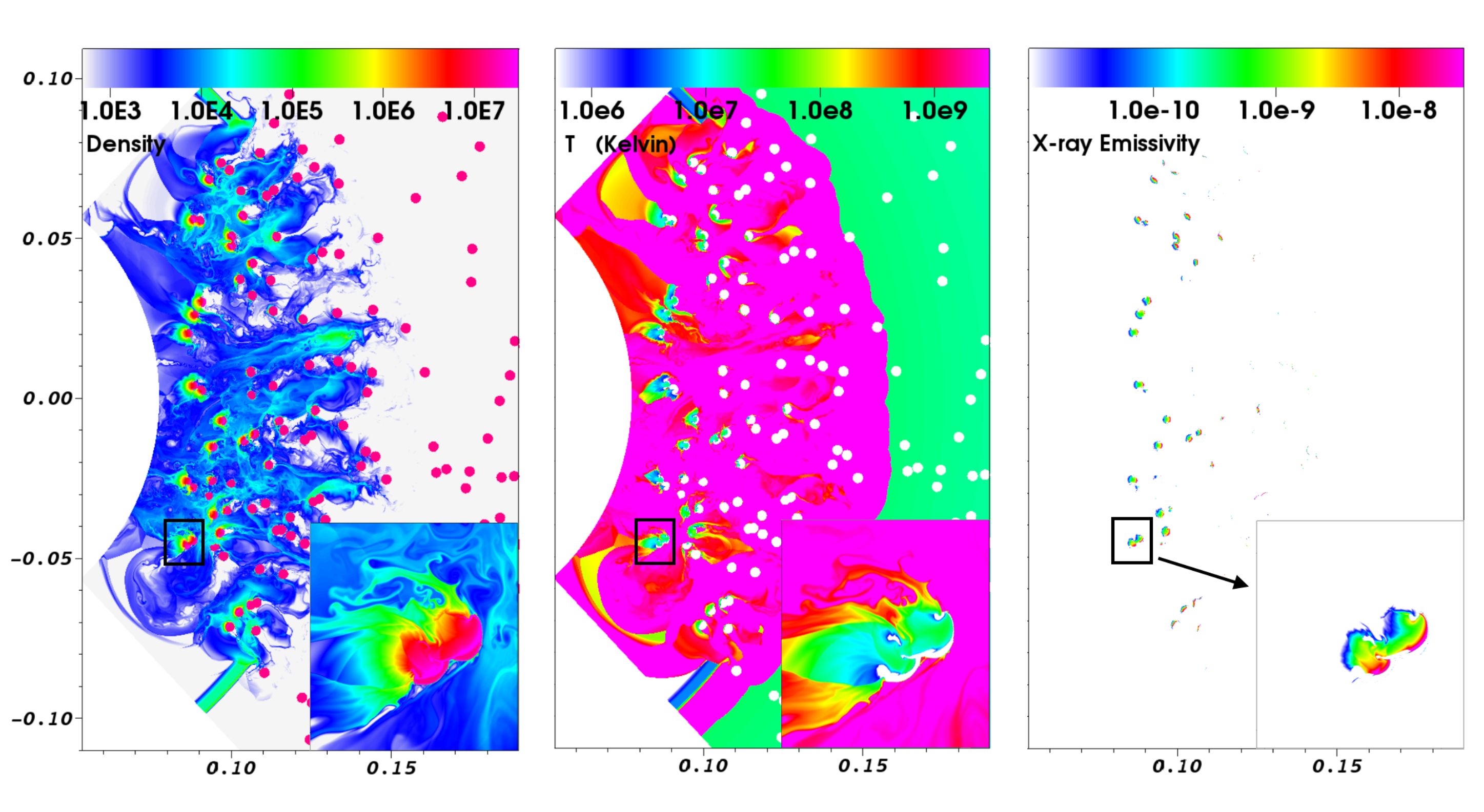}  %{SP3B.pdf} 
   \caption{Simulation setup and results for J0952. 
Top left: the initial clumpy torus for J0952. The clouds making up the torus are randomly distributed within a circle centered at 0.26 pc from SMBH with radius of 0.20 pc (blue dashed circle). The inner edge of the torus is 0.06 pc from the central BH. In our simulations, the outer boundary is smaller (black dashed line), which is enough for fitting current X-ray data. Top right: distribution of column density (Log $N_{\rm H}/{\rm cm^{-2}}$) over view angle -- $\theta$ for the simulation domain. The covering factor for $N_{\rm H}> 1\times 10^{22}$ is 54\%.  
Middle panels show the snapshots of model J1 at $\trest=4.5$ yr after TDE outburst (in the source rest frame). The left, middle and right panel show the distribution of density in units of $\mhcm$, temperature in Kelvin, and the X-ray volume emissivity $j_{\rm x}=n_{\rm e}n_{\rm H}\epsilon(T)$ in erg cm$^{-3}$ s$^{-1}$, respectively. The small window at the corner in each picture shows the enlarge view of the black frame. The materials just swept up by shocks with $10^{7-8}$ K dominate the X-ray emissions. Bottom panels show the snapshots for PS1-10adi at 4.05 yr (model PS). } 
%\vspace{0.8cm}
\label{plot3}
\end{figure*}

\begin{figure*}[!htb]
 \includegraphics[width=0.98\textwidth]{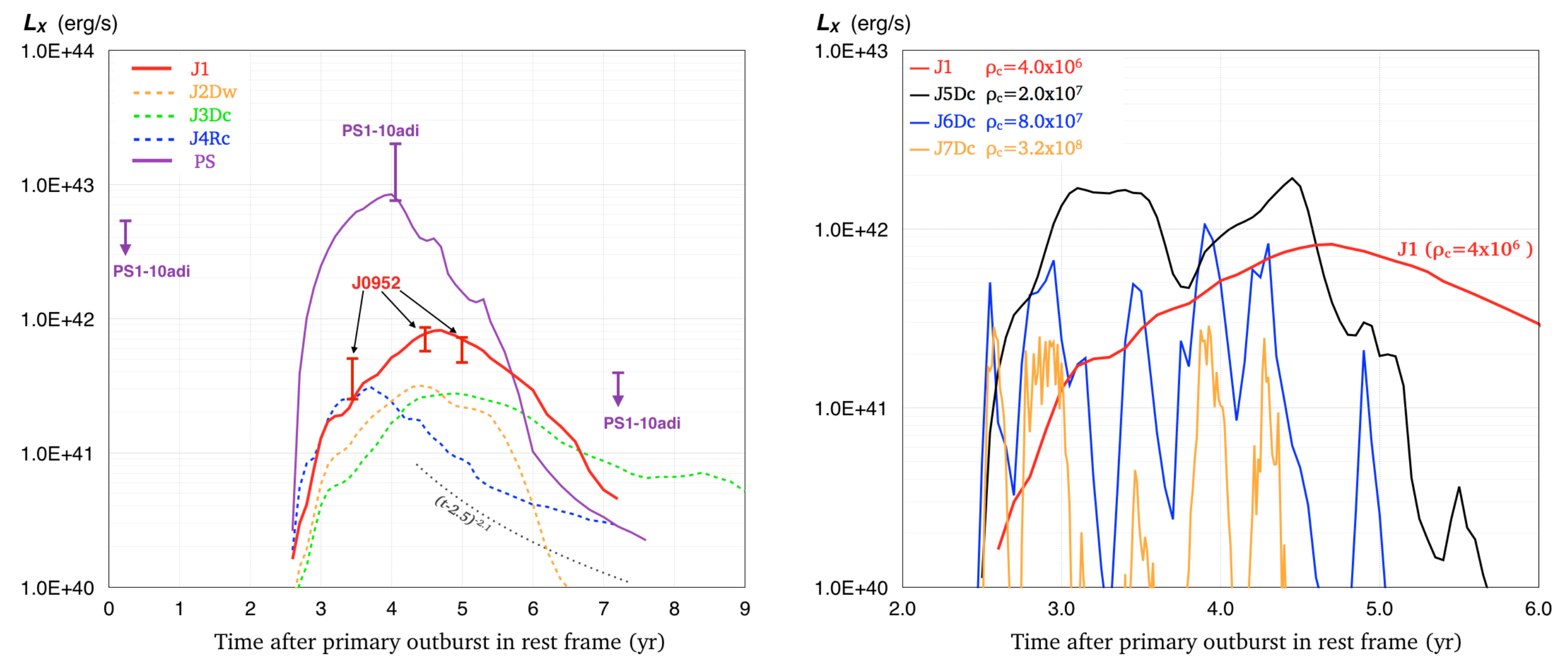}
 \caption{ 
Left panel shows the simulated X-ray light curves for J0952 in colored lines and PS1-10adi and the observational data in errorbars. The observational values are intrinsic ones (before absorptions, see appendix for details).  
The simulated intrinsic X-ray luminosity (0.3-10 keV) is calculated by the density and temperature distributions from hydrodynamic simulations. 
The abscissa is the time in the source rest frame, with the starting time representing the TDE primary outburst moment. The best fit models are plotted in solid lines (model J1 and PS).  
In the right panel, we focus on the influence of cloud density on the results. J5Dc can be roughly regarded as the transition case between strong RC mode and weak RC mode. For strong RC mode (J6Dc and J7Dc), as the cloud density increases, the X-ray luminosity decreases and the variability timescale shortens.  
} 
%\vspace{0.2cm}
   \label{plot4} 
\end{figure*}

\begin{figure*}[!htb]
\centering
   \includegraphics[width=0.46\textwidth]{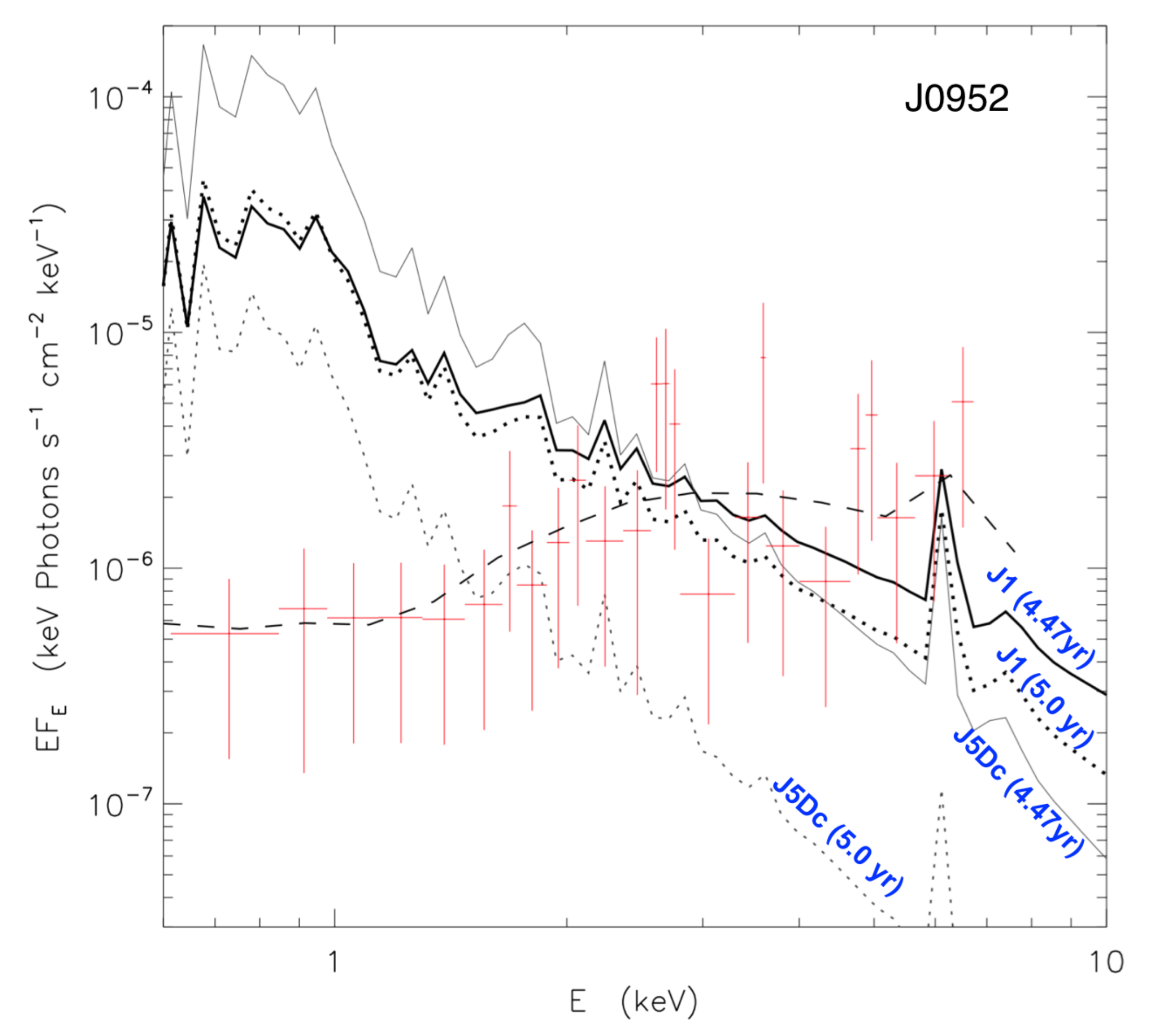} 
   \caption{ 
Comparison of X-ray spectra from theoretical models and observations (red errorbars) for J0952. The observational spectrum is stacked with three detections (observed at at $\trest=$3.45 yr, 4.47 yr and 5.0 yr). The solid and dotted lines show the unabsorbed spectra at $\trest=4.47$ yr (solid) and 5.0 yr (dotted) for model J1 (thick) and J3Dc (thin), respectively. In hard X-ray band, the spectrum of model J1 is roughly consistent with observations. 
In addition, the X-ray light curve and the evolution of the spectrum are very sensitive to the the cloud density. When the cloud density is increased by several times, the spectrum becomes softer and changes more rapidly. 
\emph{The mismatch in the low-energy range is due to that the intrinsic absorption in simulations is not included. }
For reference, using the APEC model in XSPEC, we also plot a best-fit model (dashed line) from a plasma of fixed temperature of 8 keV, which is partially absorbed by an intrinsic column density of $3.4\times 10^{22}\cms$ and a covering factor of 0.9.
We note that both the plasma temperature and the absorber's column density required are consistent with our physical model.  
The line width of the predicted emission lines given by simulations is spurious, which is due to the coarse energy resolution we used. }
%\vspace{0.2cm}
   \label{plot5} 
\end{figure*}

%%%%%%%%%%%%%%%%%%%%%%%%%%%%%%%%%%%

\section{Discussion}

It is commonly thought that the X-rays detected years after the TDE outburst may come from the newly exposed inner part of the accretion disk (e.g., \citealt{auchettl2017, jonker2020}). 
However, as long as a TDE occurred in a SMBH with dusty torus does have strong winds, the X-ray afterglow will almost inevitably arise. 
Therefore the two scenarios (disk or afterglow) may be applicable to different candidates, of which some characteristics can help us distinguish them.  
The first one is the difference in variability timescale. Drastic variability of disk X-rays could occur in hours, while for the X-ray afterglow, significant variability typically occurs in months to several years. 
We also note that for afterglow scenario, when the cloud density is very high (e.g., $\ga 10^8\mhcm$), the variability timescale can  also be as short as weeks or days, but the intrinsic X-ray luminosity is difficult to exceed $10^{42}~\ergs$. 
The second one is the existence of shocks. There must be strong shocks in the X-ray afterglow model, which will accelerate particles and generate radio emission, $\gamma-$rays and neutrinos. X-ray radiation from disk is not necessarily accompanied by these phenomena, which if detected, must resort to jet.  

The unbound debris that escaped during the initial disruption may also impact the torus, with a typical velocity of a few thousand $\kms$ and kinetic energy of $\sim 10^{50}$ erg (but see also \citealt{guillochon2016}). The unbound debris is extremely elongated in the radial direction which must connect to the bound stream, and is quite narrow and thin in the plane perpendicular to the moving direction. Its solid angle viewed from the SMBH is $\Omega_{\rm ud} \sim \beta (M_{\rm BH}/M_{\rm star})^{-1/2}$ (\citealt{kasen2010}), where $\beta$ is the impact factor defined as the ratio of the tidal radius to the distance from the pericenter to the SMBH.  
If the self-gravity of the debris is included, part of the unbound debris would be confined by self-gravity (\citealt{steinberg2019, coughlin2016}), and the solid angle is expected to be smaller. 
For a mass ratio $M_{\rm BH}/M_{\rm star}$ of $10^{6}$ and $\beta \simeq 1$, the solid angle of the debris is typically $\Omega_{\rm ud} \sim 1\times 10^{-3}$ sr. Due to the extremely small solid angle, the X-ray generation region is much smaller than wind-torus interactions, and the expected X-ray luminosity would be much lower than the latter case ($\sim 10^{39}\ergs$ according to our simulations; Mou et al. in preparation).

According to above analysis and numerical simulations, we argue that TDEs occurred in SMBHs with dusty tori are a special category, in which the predicted X-ray afterglow provides rich and valuable information.  
Firstly, it provides an excellent opportunity to ``dissect'' the torus clouds, and constrain its physical parameters. 
Secondly, it can be used to make constraints on wind parameters including the energy and mass, and furthermore, the lower limit of the mass of disrupted star. 
Thirdly, it provides a new method for the ex post verification of TDE events years after the TDE outburst.
Accordingly, we propose a ``diagnosis'' scheme on exploring TDE and torus by the X-ray afterglow (Figure 6).  

\begin{itemize}

\item An X-ray afterglow should be associated with an earlier \emph{infrared echo}. The time lag between the X-ray afterglow and optical outburst reveals the velocity of TDE winds $\vw$, in which the distance of the inner radius of dusty torus can be inferred from observations. 
Once $\vw$ is known, there are two remaining unknown parameters: $\rhow$ and $\rhoc$. This provides the 1st restriction on the parameters of wind and torus (here $\vw$ only).   

\item If the X-ray spectrum can be obtained, the temperature range can be inferred. This provides the 2nd independent restriction on the parameters of wind and torus (here is the density contrast $\rhoc/\rhow$, as $\vw$ is known). 

\item The X-ray luminosity can be as high as $\sim 10^{41}~\ergs$ or even higher, which provides the 3rd independent restriction on the parameters of wind and torus (here are $\rhoc$ and $\rhow$). 
\end{itemize}

The above three observable parameters (time lag, spectrum and luminosity) are enough to constrain the basic parameters: $\vw$, $\rhow$ and $\rhoc$. 
In addition, the plateau timescale $t_{\rm plat}$ reveals the cooling timescale of the shocked cloud $\tc$ or the timescale of the cloud shock sweeping over one cloud $\tsc$, which also roughly applies to the winds + multi-clouds interactions. Specifically, $t_{\rm plat}$ is comparable to $\tc$ for strong RC mode, while it is comparable to $\tsc$ for weak RC mode. In the former case, this provides another independent condition for testing the above three parameters, while for the latter case, it can be used to estimate the cloud size $\Rc$. 
%{\bf The rate at which the luminosity reaches the peak and enters the power-law decay also indicates the cloud size. }
Besides, the decay rate of the X-ray light curve is sensitive to the cloud density (model J1, J3Dc and 
J5Dc-J7Dc 
in Figure 4). As a redundant condition, this still can be used to check the above results.  
Finally, it should be emphasized that the parameters of clouds and TDE winds in our model are simplified. In practice, the dusty torus should be composed of clouds of various sizes, densities and the distribution of the clouds is also uncertain. The winds actually should be anisotropic, of which the energy/mass should be mainly concentrated within a certain angle range. Considering the complexity, the parameters inferred by our scheme actually sketch the approximate values. More observational data will help reduce the degrees of freedom of the parameters, and provide more reliable constraints.  

\begin{figure*}[!htb]
   \centering
   \includegraphics[width=0.8\textwidth]{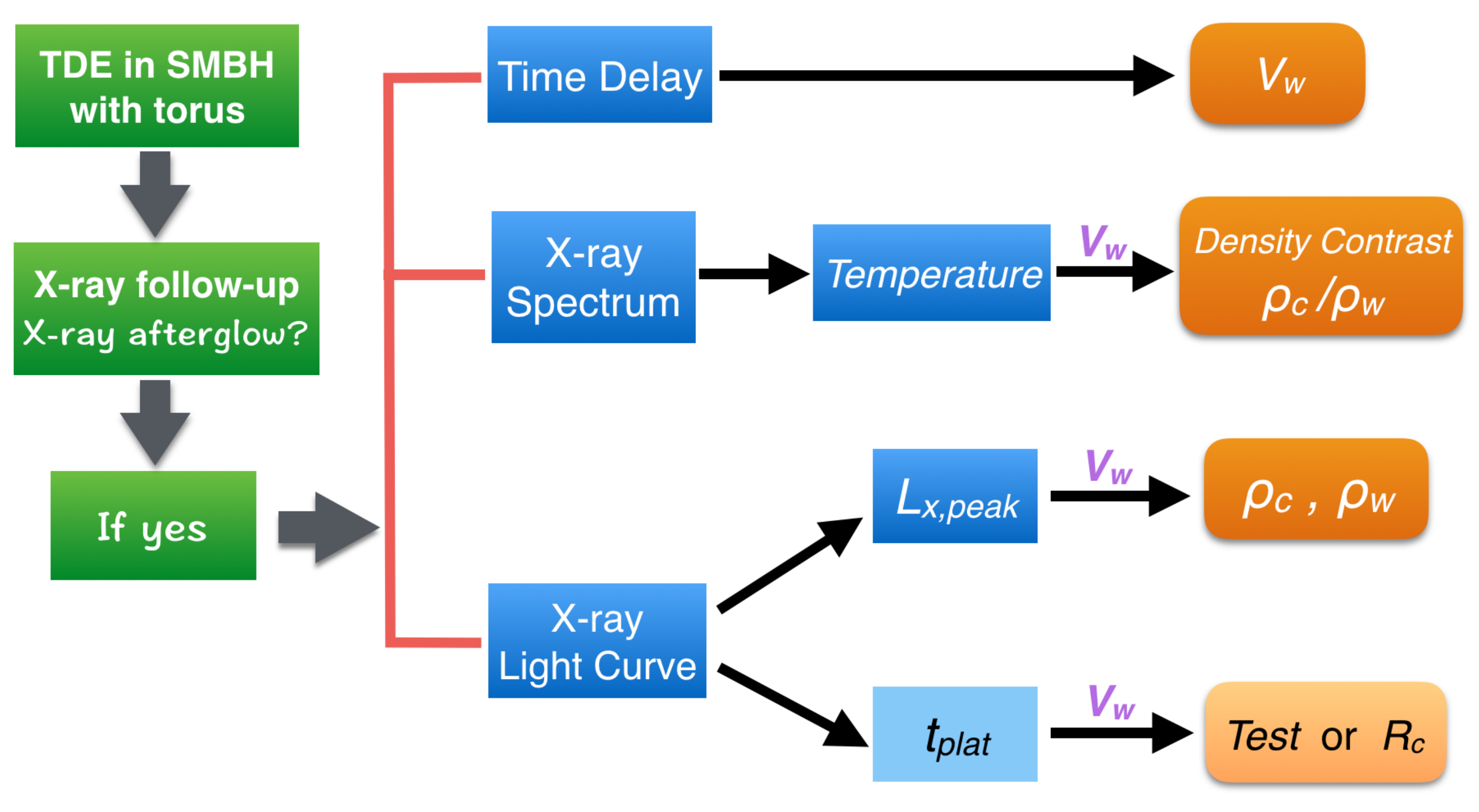} 
   \caption{The ``diagnosis'' scheme on exploring TDE and torus by TDE wind. The basic physical parameters include the wind velocity $\vw$, wind density $\rhow$ and cloud density $\rhoc$ at the inner edge of the torus. The X-ray follow-up provides three independent observations: the time delay of the X-ray afterglow, the X-ray spectrum and the X-ray peak luminosity, which are enough to constrain those three parameters. The duration of the X-ray plateau provides another independent condition for testing those parameters or restricting the size of the cloud. In addition, the decline rate of the X-ray light curve is sensitive to the cloud density, which can also be used to test the above results. } 
   \label{plot6}
\end{figure*}

%%%%%%%%%%%%%%%%%%
%%%%%%%%%%%%%%%%%%

\acknowledgements
We acknowledge financial support by National Science Foundation of China (NSFC-11703022, 11833007, U1731104, 11733001, 12073025, 11873072, 11822301). 
G.M. is supported by ``the Fundamental Research Funds for the Central Universities'' (No. 2042019kf0040) and the opening fund of the Key Laboratory of Galaxy Cosmology, Chinese Academy of Sciences (No. 18010202). 
T.W. is supported by NSFC through grant NSFC-11833007 and 11421303.  
F.G. thanks the support by Natural Science Foundation of Shanghai (No. 18ZR1447100) and Chinese Academy of Sciences (No. QYZDB-SSW-SYS033). 
W.W. is supported by the National Program on Key Research and Development Project (Grants No. 2016YFA0400803) and the NSFC (11622326 and U1838103). 
Z.H. is supported by NSFC-11903031 and USTC Research Funds of the Double First-Class Initiative YD 3440002001.

%\bibliography{ref}

%%%%%%%%%%%%%%%%%%%%%%%%%%%
%%%%%%%%%%%%%%%%%%%%%%%%%%%

%\clearpage

%%%%%%%%%%%%%%%%%%%%%%%%%%%
%%%%%%%%%%%%%%%%%%%%%%%%%%%

\appendix

\noindent \textbf{Wind + One Cloud Instantaneous Interaction}

To investigate the wind+one cloud instantaneous interaction, we have performed hydrodynamic simulations. 
We consider a simple case here: short-term wind collides with one circular cloud located at $\theta=90^{\circ}$ in spherical coordinates (2.5 dimensional simulations with $\partial_{\phi} \equiv 0$,  see the main text for the fluid equations, the codes and simulation setup). 
The isotropic wind lasts for $\dtbst=100$ days (or 25 days for model \emph{ATbst} and \emph{CTbst}), with velocity of 0.1c. The total wind mass is fixed at 0.78$\msun$ and the kinetic energy is $7.0\times 10^{51}$ erg. The density of the wind is $3\times 10^4 \mhcm/r^2_{0.1}$ (or $1.2\times 10^5 \mhcm/r^2_{0.1}$ for model \emph{ATbst} and \emph{CTbst}). The initial cloud is located at 0.1 pc from the SMBH, with radius $\Rc$ of $1.6\times 10^{-3}$ pc ($5\times 10^{15}$ cm). The initial cloud densities are tested in four cases: uniform densities of $2.4\times10^7$, $6\times10^6$, $1.5\times 10^6~\mhcm$, and non-uniform density of $2\times 10^7~\mhcm exp(-2r_c/\Rc)$ where $r_c$ is the distance to the cloud center. 
The parameters of simulation setup and results for simulation tests are shown in Table 3, and we plot the evolution of the internal energy of shocked cloud and the X-ray light curves in Figure 7. 

In pure adiabatic case without radiative cooling (RC), the total internal energy of the cloud will undergo three stages: 
S1, wind-cloud interaction stage;  
S2, cloud shock propagating stage without external wind; 
S3, adiabatic expansion stage after the cloud shock sweeping across the cloud. 

When the RC is included, the evolution of $E_{\rm c}$ and X-ray luminosity (0.3--10 keV band) can be classified into two modes: strong RC mode in which $E_{\rm c}$ and the X-ray luminosity evolve drastically within months (model \emph{CD27}) , and weak RC mode in which $E_{\rm c}$ and X-ray luminosity evolve relatively slowly (model \emph{CD66, CD16, CDnu, CTbst}). In addition, by comparing models \emph{CD66} and \emph{CTbst}, we find that when the mass and energy of the wind are fixed, the X-ray light curves are not sensitive to the duration of the wind $\dtbst$.  

For the energy conversion efficiency $\eta_{E}$ in pure adiabatic process, we also find that the expression $0.56\cv \chi^{-0.5}$ is consistent with simulation results. 

\begin{table}
  \centering
  \renewcommand{\thefootnote}{\thempfootnote}
  \caption{Parameters of simulation setup and results for wind+one cloud instantaneous interactions: (1) model label, (2) whether including radiative cooling (RC) or not, (3) cloud density, (4) duration of the wind, (5) duration of the X-ray plateau (above $0.5\lxpeak$), (6) decay time of the X-ray (from 0.5$\lxpeak$ to 0.1$\lxpeak$), (7) the peak X-ray luminosity, (8) maximum internal energy of the cloud during S2, (9) energy ratio of $E_{\rm c,max}$ over $E_{\rm wind}$, (10) $\eta_{E}=0.56\cv \chi^{-0.5}$. }
  \begin{tabular}{ c  c  c  c  c  c  c  c  c  c}
  \hline
         {Model}
        & {RC}
        & {$\rhoc$}
        & {$\dtbst$}
        & {$t_{\rm plat}$ }
        & {$t_{\rm decay}$}
        & {$\lxpeak$}
        & {$E_{\rm c,max}$}
        & {$E_{\rm c,max}/E_{\rm wind}$}
        & {$\eta_{E}$}
        \\
      & (Yes/No) &($\mhcm$)  & (day) & (year)   &  (year)  & $\ergs$ & erg  &   &   \\
    (1)   &  (2)        &     (3)      & (4)           &  (5)            &  (6)  &  (7)  & (8)  & (9)  & (10) \\
    \hline
    AD66 & N & $6 \times 10^6$ & 100 & --  & --  & -- & $5.0\times 10^{48}$ & $7.2\times 10^{-4}$ & $6.4\times 10^{-4}$\\ % SP76
    CD66 & Y & $6 \times 10^6$ & 100 & 1.6  & 0.5  & $1.05\times 10^{41}$ & $4.2\times 10^{48}$ & $6.0\times 10^{-4}$ & $6.4\times 10^{-4}$ \\ % SP77
    AD27 & N & $2.4\times 10^7$ & 100 & -- & --  & -- &$2.7\times 10^{48}$ & $3.8\times 10^{-4}$ & $3.2\times 10^{-4}$\\ % SP78
    CD27 & Y & $2.4 \times 10^7$&100 &0.15 &$<0.05$ & $2.68\times 10^{41}$  & $1.1\times 10^{48}$ & $1.5\times 10^{-4}$ & $3.2\times 10^{-4}$\\ % SP79
    AD16 & N & $1.5 \times 10^6$ & 100 & -- & --  & -- &$7.8\times 10^{48}$ & $1.1\times 10^{-3}$  & $1.3\times10^{-3}$ \\ % SP82
    CD16 & Y & $1.5 \times 10^6$ & 100 & 0.8 & 0.9 & $1.5\times 10^{40}$ & $7.5\times 10^{48}$ & $1.1\times 10^{-3}$ & $1.3\times10^{-3}$  \\  % SP83
    ADnu & N & (0.27-2)$\times 10^7$&100& -- &--&--& $6.1\times 10^{48}$ & $8.8\times 10^{-4}$ & $6.5\times 10^{-4}$\\  % SP84
    CDnu & Y & (0.27-2)$\times 10^7$&100& 0.4& 0.1 &$2.13\times 10^{41}$& $5.4\times 10^{48}$ & $7.8\times 10^{-4}$ & $6.5\times 10^{-4}$ \\  % SP85
    ATbst & N &$6 \times 10^6$ & 25 &  --  & -- & --& $1.16\times 10^{49}$ & $1.66\times 10^{-3}$ & $1.3\times 10^{-3}$ \\ % SP377
    CTbst & Y &$6 \times 10^6$ & 25 & 1.3  & 0.9 & $9.3\times 10^{40}$ & $1.12\times 10^{49}$ & $1.60\times 10^{-3}$ & $1.3\times 10^{-3}$ \\ % SP376
    \hline
 \label{table3}
  \end{tabular}
\end{table}

\noindent \textbf{Fitting the X-ray Emissivity $\epsilon(T)$ }

We plot both the X-ray emissivity data from APEC and the fitting curve (described by Equation 9 and 10) in the left panel in Figure 8.  
The analytical expressions (solid line) can fit the data well.

\noindent \textbf{Tests on Numerical Resolutions} 

The fiducial resolution is adopted in CD66 (Table 3), where the size of the mesh at $r=0.1$ pc is $2.2\times 10^{-5}$pc in $r-$direction and $2.3\times 10^{-5}$ pc in $\theta-$direction. Therefore, the radius of the cloud is resolved by $\sim 70$ meshes. 
We tested a high resolution case (doubled resolution marked as ``0.5dX''), a low resolution case (doubling the grid size, marked as ``2.0dX''), and a very low resolution case (4 times the grid size, marked as ``4.0dX''). The results are plotted in the right panel in Figure 8, from which we conclude that the results are convergent on resolutions, and the fiducial resolution is sufficient to ensure the reliability of the results. For simulations in the main text, the resolutions are close to the fiducial one here. Specifically, the grid size at $r=0.1$ pc is $(dr, r d\theta)$=$(2.2\times 10^{-5} {\rm pc}, 2.7\times 10^{-5} {\rm pc})$ for J0952 and $(1.8\times 10^{-5} {\rm pc}, 2.3\times 10^{-5} {\rm pc})$ for PS1-10adi.

 \begin{figure}
   \centering
   \includegraphics[width=0.9\textwidth]{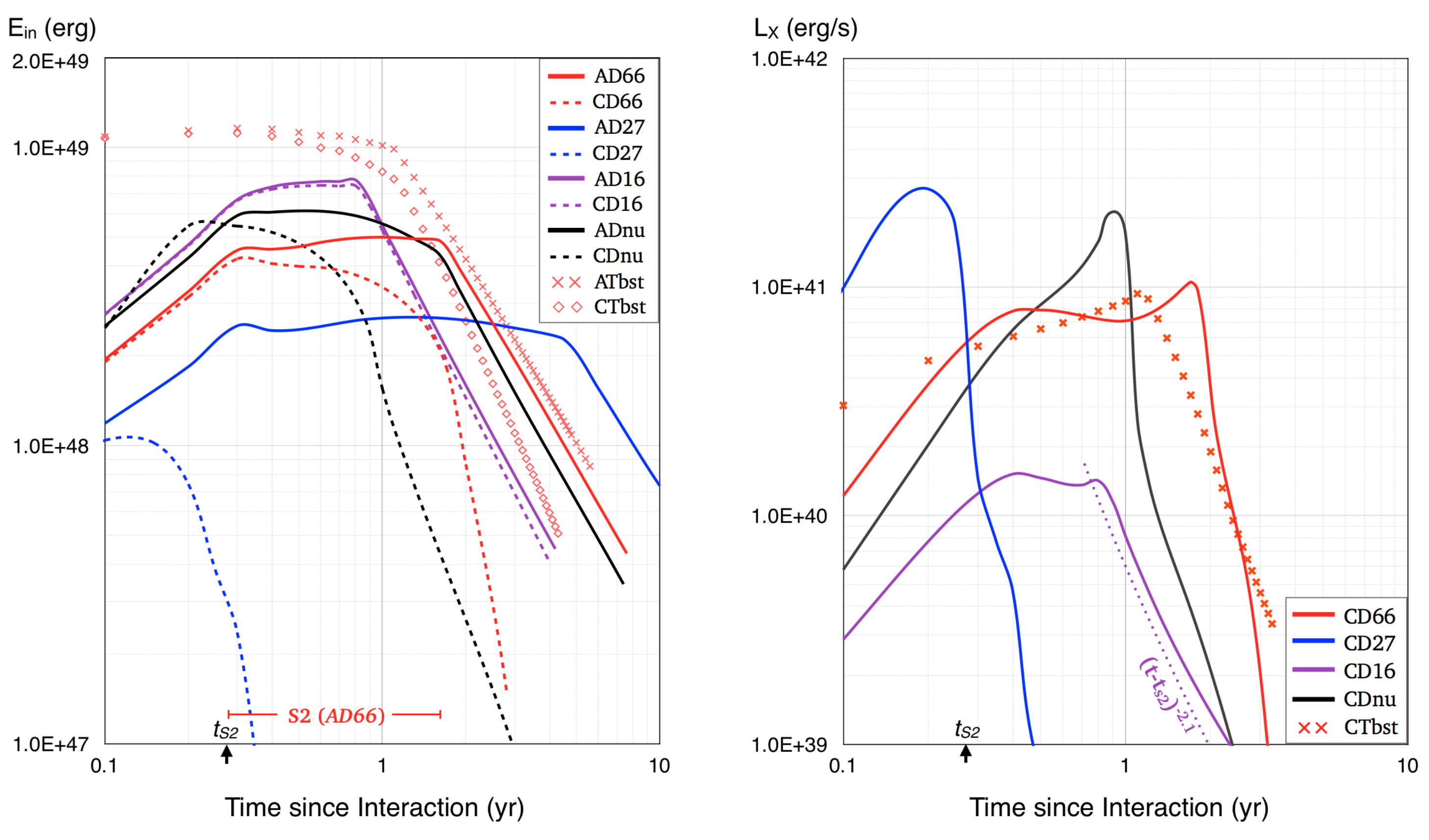}  %{SP1iv.pdf} %{SP1EL.pdf} 
   \caption{ Left: evolution of the internal energy of shocked cloud ($E_{\rm c}$) for cases without radiative cooling (solid lines and red crosses, model labels beginning with \emph{A}), and with radiative cooling (dashed lines and red diamonds, model labels beginning with \emph{C}). 
For the pure adiabatic case, the evolution of $E_{\rm c}$ is obviously can be divided into three stages. During S1 when the wind is interacting with the cloud, $E_{\rm c}$ sharply increases with time. During S2, $E_{\rm c}$ varies little, which also holds for the cloud with non-uniform density (model ADnu). Here we specifically marked the S2 for AD66. During S3, the cloud starts to expand efficiently and $E_{\rm c}$ decrease as a power-law function of time in $(t-\tst)^{-m}$, in which the index $m \simeq 1.4$. 
Right: two modes of X-ray light curve. For strong RC (blue solid line), the X-ray luminosity changes drastically within several months. For weak RC (the other lines), X-ray changes relatively slowly. See Table 3 for more details on the timescales. }  
\label{plots7}
\end{figure}

\begin{figure}[!htb]
   \centering
   \includegraphics[width=0.45 \textwidth]{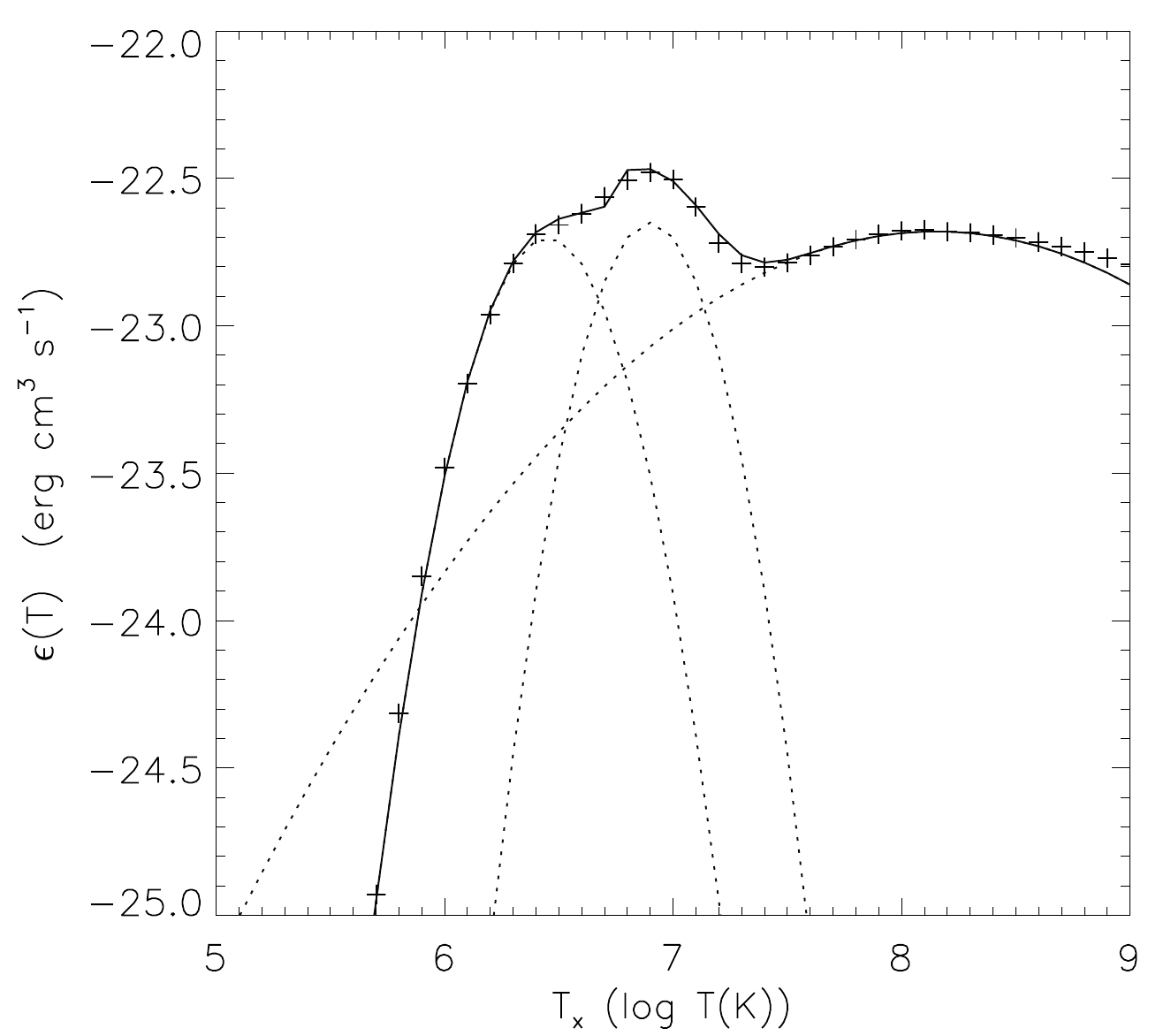}  \hspace{0.3cm}
   \includegraphics[width=0.45 \textwidth]{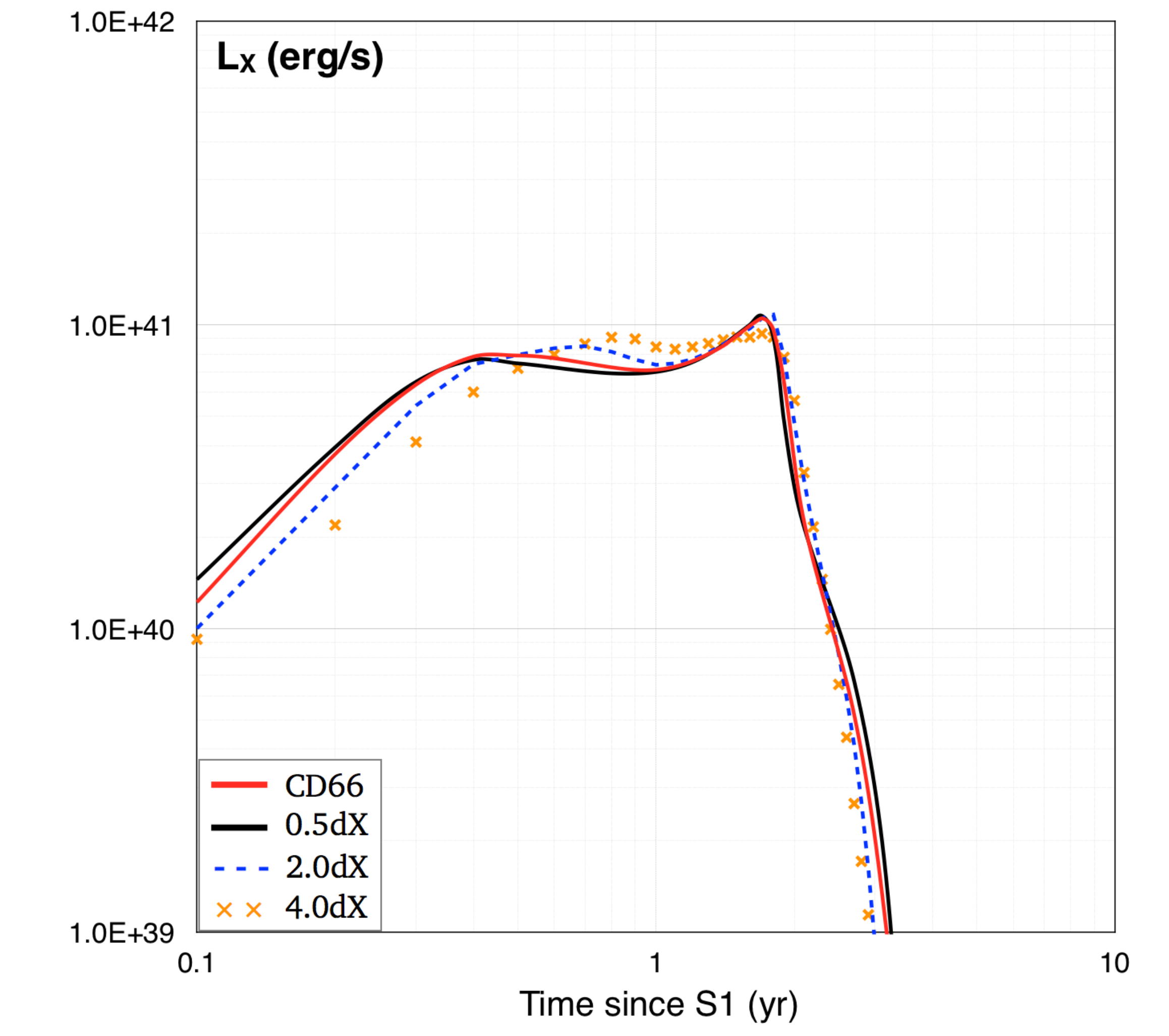} 
%   \vspace{-2.0cm}
   \caption{Left: we fit the X-ray emissivity data in 0.3--10 keV band from APEC (crosses) with analytical expressions (Equation 9-10, solid line), and the dotted lines represent the three components in the right side of Equation 10. Right: evolutions of X-ray luminosity under different numerical resolutions. We tested four resolutions: the fiducial resolution marked as CD66, 0.5$\times$, 2.0$\times$, and 4.0$\times$ fiducial one. The results are convergent on the numerical resolutions, and fiducial resolution is enough to ensure the reliability of the result. } 
   \label{plots8}
\end{figure}

%%%%%%%%%%%%%%%%%%%%%%%%
%%%%%%%%%%%%%%%%%%%%%%%%

\noindent \textbf{X-ray Data}

 {\bf  Chandra observations} 
J0952 was observed 3 times by Chandra/ACIS-S observation (Table 4). The Chandra data reductions and the spectra are done following the standard procedures with the  Software of CIAO v.4.12 with the latest calibrations v.4.9.1.  In order to increase the S/N ratio and get the spectral profile, a stacking spectrum is made for J0952, with the 3 observations.

{\bf Swift observations}
PS1-10adi was observed 7 times by {\it Swift/XRT} observation after discovered its optical flare.
Four observations are in 2010 with total exposure of 15.2 ksec, and three observations are in 2015 with total exposure of 7.0 ksec.  
J0952 was observed 7 times by {\it Swift/XRT} observation in 2015 with total exposure of 9.9 ksec.
We download the data from the  UK Swift Science Data Centre, and reduce the {\it Swift/XRT} Observations with the software HEASoft (V.6.26)  and the latest updated calibration files of {\it Swift}.  We reprocess the event files with the task `xrtpipeline', and select the event files which operated in Photon Counting mode. 
The source file is extracted using a source region of radius of 47.1'' (20-pixel). The background is estimated in an annulus source-free region centered on the source position.  
However, the two sources are too faint for {\it Swift/XRT} observation. We finally stack the events for J0952 together, though it is still not detected. 
The events of PS1-10adi are also stacked into two groups. One group includes the four observations in 2010, the other group includes the observations in 2015.
We find that it is not detected in X-ray band in 2010, even after stacking the events. 11 net photons are detected in 0.3--10 keV in 2015. The count rate is $5.0\pm3.3\times10^{-4}$, $8.9\pm4.1\times10^{-4}$, and $1.61\pm0.57\times10^{-3}$  cts s$^{-1}$, respectively, in 0.3-2, 2-10, and 0.3-10 keV. 

{\bf XMM-Newton observations}
PS1-10adi was also observed by XMM-Newton observation in 2019-5-20. We download the data from 
the XMM-Newton Science Archive data centre, and reduce data with the Software SAS (v.18) with the most updated calibration files. Following the standard SAS procedures, the cleaned events file, source region is created in. However, no X-ray photons is detected in. We estimate the upper count rate from the data from pn detector.  

{\bf Spectrum}
All the (stacking) spectra were grouped to have least 2 counts in each bin, if the source is detected. 
The X-ray spectral fitting is performed using XSPEC (v.12.11) and adopting C-statistic. A simple power-law model including the Galactic absorption ($N_{H} =6.51\times10^{20} \cms$ for PS1-10adi, $N_{H} =2.39\times10^{20} \cms$ for J0952, \citealt{bekhti2016}) is applied to the spectrum. 

For PS1-10adi, we get the best-fit photon index $\Gamma=0.91\pm0.88$ (C-Statistic/d.o.f=2.47/5). If fixed the photon index at the typical AGN value of $\Gamma=2$, the C-Statistic/d.o.f would be 6.39/6.  
For J0952, we get the best-fit photon index $\Gamma=0.24\pm0.40$ (C-Statistic/d.o.f=18.79/23). 
We note that the X-ray spectra for both sources are rather hard, especially compared with thermal X-ray bright TDEs or typical AGNs. Thus, the results may indicate the sources are (partially) obscured. We then apply a partial absorbed APEC model to fit the spectrum of J0952 with a plasma at fixed temperature of 8 keV, and obtain an intrinsic absorber with $N_{H} = 3.4^{+3.3}_{-1.8} \times 10^{22} \cms$, and covering factor of $f=0.89^{+0.06}_{-0.10}$ (C-Statistic/d.o.f=17.31/22). 
Our fitting result is shown in Figure 5 (dashed lines).  

The observation logs and results are listed in Table 4. The upper limit count rates for Swift observations are also estimated. The flux and luminosity in 0.3--10 keV is estimated from the best-fit PL model for the stacking spectrum.

\begin{table}[htp]
\caption{X-ray observation logs.
Col. (1), observation date; Col. (2), observation ID; Col. (3) net exposure time in unit of ksec; Col. (4),  X-ray count rate in 0.3--10 keV band; Col. 5 and Col. (6), the estimated 0.3-10 keV band flux and luminosity from best-fit model after corrected the absorption.}
\begin{center}
\begin{tabular}{|cccccc}
(1) & (2) & (3) & (4) & (5) & (6) \\
date & obsid & exposure time & count rate & flux & luminosity \\
        &             &   ksec             & $10^{-3} {\rm cts/s}$    & $10^{-14} {\rm erg ~cm^2 ~s^{-1}}$ & $10^{42} {\rm erg ~s^{-1}}$ \\

\multicolumn{6}{c}{\bf J0952} \\
\multicolumn{6}{c}{\bf Chandra/ACIS-S} \\  %\end{tabular}
2008-02-05 & 9814         &  9.8  & $0.81\pm0.29$ &   $1.9^{+0.5}_{-0.7}$  &  $0.40^{+0.10}_{-0.15}$  \\
2009-03-15 & 10727        &  16.7 & $1.52\pm0.32$ &  $3.6^{+0.6}_{-0.7}$  &   $0.72^{+0.13}_{-0.15}$  \\
2009-10-06 & 10728        &  16.9 & $1.43\pm0.30$ &  $3.0^{+0.5}_{-0.7}$  &   $0.62^{+0.10}_{-0.15}$  \\
\multicolumn{6}{c}{\bf Swift/XRT} \\  %\end{tabular}
2015-4-12 &00092115001  & 0.54 & \nodata & \nodata & \nodata \\
2015-4-15 &00092115002  & 0.20 & \nodata & \nodata & \nodata \\
2015-4-16 &00092115004  & 1.33 & \nodata & \nodata & \nodata \\
2015-4-17 &00092115005  & 0.24 & \nodata & \nodata & \nodata \\
2015-4-20 &00092115006  & 0.31 & \nodata & \nodata & \nodata \\
2015-4-21 &00092115007  & 0.33 & \nodata & \nodata & \nodata \\
2015-6-23 &00092115008  & 6.83 & \nodata & \nodata & \nodata \\
2015-04 - 2015-06          & stacking 1-8  & 9.9  & $<0.5$    &  $<4.4$    &    $<0.9$            \\

\hline
\multicolumn{6}{c}{\bf PS1-10adi} \\
\multicolumn{6}{c}{\bf Swift/XRT} \\
2010-10-06 &00031834001  & 5.29 & \nodata & \nodata & \nodata \\
2010-10-07 &00031834002  & 4.24 & \nodata & \nodata & \nodata \\
2010-10-09 &00031834003  & 0.99 & \nodata & \nodata & \nodata \\
2010-12-08 &00031834004  & 4.77 & \nodata & \nodata & \nodata \\
2015-07-18 &00031834005  & 2.26 & \nodata & \nodata & \nodata \\
2015-07-19 &00031834006  & 3.45 & \nodata & \nodata & \nodata \\
2015-07-20 &00031834007  & 1.32 & \nodata & \nodata & \nodata \\
2010-end    & stacking 1-4   & 15.2 &$< 0.23$ & $ <5.4$ &  $<5.3$ \\
2015-07 	  & stacking 5-7   &  7.0  & $1.6\pm0.6$ & $16.0^{+4.4}_{-8.3}$ & $15.7^{+4.3}_{-8.2}$ \\
\multicolumn{6}{c}{\bf XMM-newton/pn} \\
2019-05-20 & 0841710101 & 21.3  &  $<0.3 $  & $<0.4$ & $<0.39$ \\

\end{tabular}

\end{center}
\label{xray}
\end{table} %

%%%%%%%%%%%%%%%%%%%
%%%%%%%%%%%%%%%%%%%

\clearpage

%%%%%%%%%%%%%%%%%%%%%%%%
%%%%%%%%%%%%%%%%%%%%%%%%

%\end{thebibliography}

\clearpage

\end{document}